\documentclass[reprint,amsmath,amssymb,aps, floatfix]{revtex4-2}
\usepackage{graphicx}
\usepackage[hidelinks]{hyperref}
\usepackage[]{units}
\usepackage{verbatim}
\usepackage{xcolor}
\usepackage[english]{babel}
\usepackage[absolute,overlay]{textpos}
\usepackage{geometry}

\newcommand{\km}{k_\mathrm{bola}}
\newcommand{\ko}{k_\mathrm{0}}
\newcommand{\kBT}{k_\mathrm{B}T}
\newcommand{\fbi}{f_\mathrm{bi}}
\newcommand{\uf}{u_\mathrm{f}}
\newcommand{\Rring}{R_\mathrm{ring}}
\newcommand{\rcross}{r_\mathrm{cross}}
\newcommand{\rmm}{r_\mathrm{m}}
\newcommand{\ts}{t_\mathrm{s}}
\newcommand{\rc}{r_\mathrm{c}}

\begin{document}

\begin{textblock*}{0.95\paperwidth}(0.5cm,0.5cm)
\centering
This article may be downloaded for personal use only. 
Any other use requires prior permission of the author and AIP Publishing. 
This article appeared in \textit{J. Chem. Phys. 14 April 2026; 164 (14): 144902} and may be found at \url{https://doi.org/10.1063/5.0325170}.
\end{textblock*}

\title{Cracking donuts and sorting lipids: Geometry controls archaeal membrane stability and lipid organization}

\author{Felix Frey$^{1,2}$}
\email{f.f.f.frey1@tue.nl}
\author{Miguel Amaral$^{1}$}
\author{An\dj ela \v{S}ari\'{c}$^{1}$}
\affiliation{$^{1}$Institute of Science and Technology Austria, Klosterneuburg, Austria}
\affiliation{$^{2}$Present address: Soft Matter \& Biological Physics, Department of Applied Physics, Eindhoven University of Technology, Eindhoven, The Netherlands}

\date{\today}
\begin{abstract}
Cells are defined by lipid membranes that differ in their structure across the tree of life.
While the membranes of most bacteria and eukaryotes consist of single-headed bilayer lipids, the membranes of archaea are composed of mixtures of single-headed bilayer lipids and double-headed bolalipids.
Archaeal bolalipids can adopt straight or u-shaped conformations, enabling them -- together with bilayer lipids -- to control whether membranes form bilayer or monolayer structures.
Yet, the physical principles governing archaeal membranes remain largely unexplored, especially how membrane structure couples to externally imposed curvature during membrane remodeling.
Here, we perform coarse-grained molecular dynamics simulations of toroidal vesicles to systematically probe the effects of all relevant combinations of mean and Gaussian curvatures on shape stability and lipid organization.
We find that soft bilayer membranes can sustain all curvatures induced, whereas rigid bolalipid monolayer membranes either transition to different vesicle shapes or rupture.
Bilayer-mimicking u-shaped bolalipids and bilayer lipids are spatially accumulated in regions of high mean membrane curvature independent of Gaussian curvature.
Our work identifies curvature-composition coupling as a physical signature  of archaeal membrane remodeling.
\end{abstract}

\maketitle

\begin{figure*}
\centering
\includegraphics[width=.95\textwidth]{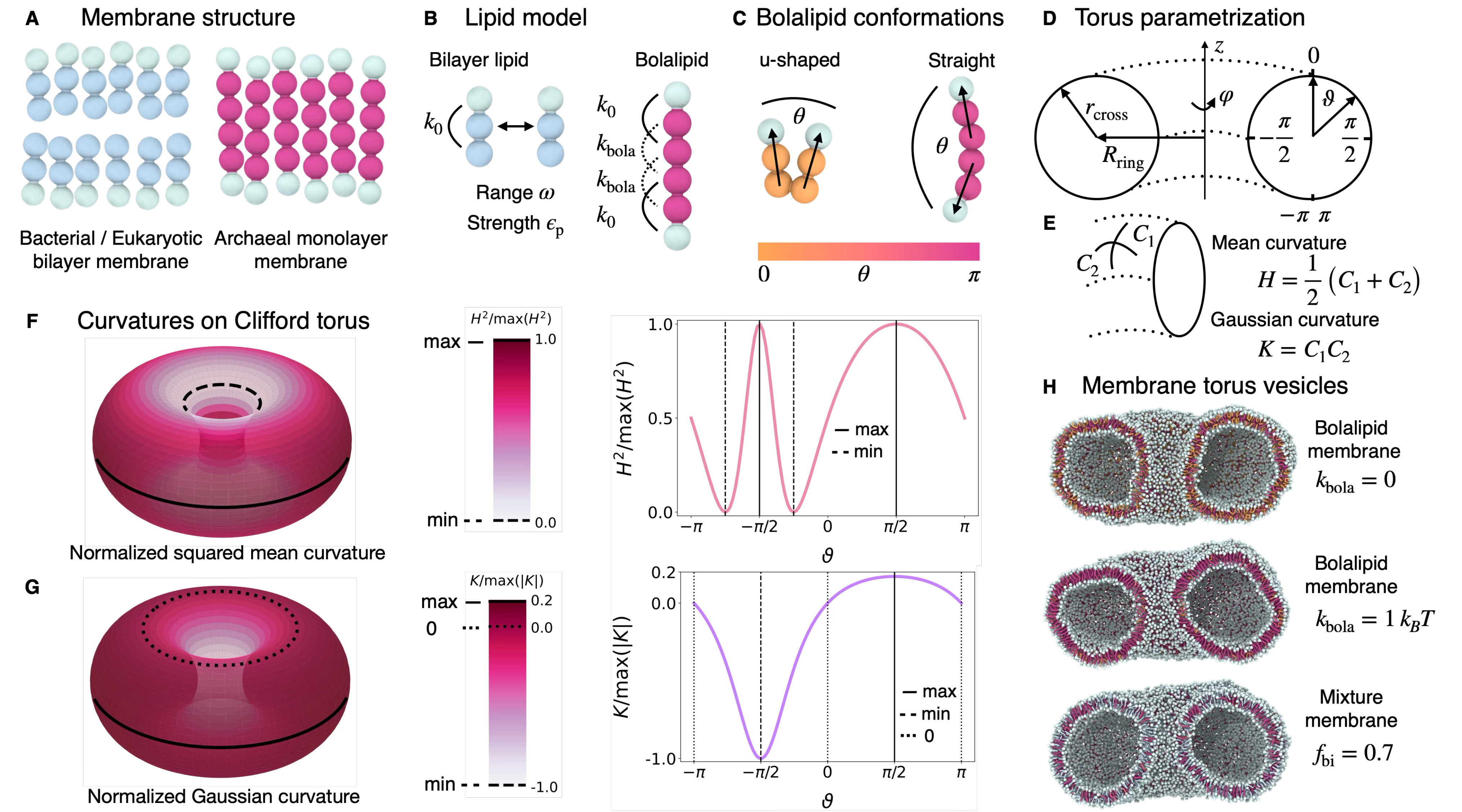}
\caption{Computational model, toroidal vesicle geometry and curvature distribution on toroidal vesicles.
(A, left) Bacterial or eukaryotic bilayer membrane. 
(right) Archaeal monolayer membrane. 
(B, left) A bilayer lipid consisting of 1 head bead (gray) and 2 tail beads (cyan). 
The molecular stiffness is controlled via $k_0$.
(right) An archaeal bolalipid consisting of 2 head beads (gray) and 4 tail beads (pink).
The molecular stiffness is controlled via $k_0$ and $\km$.
Bilayer and bolalipid tail beads attract each other, with $\omega$ and $\epsilon_\mathrm{p}$ controlling the range and strength of the hydrophobic interaction along lipid tail beads. 
(C) Bolalipids adopt two dominating conformations: u-shaped (orange) and straight shape (pink).
(D) Torus geometry with ring radius $\Rring$ and cross section radius $\rcross$. 
The toroidal angle $\vartheta$ parametrizes the cross-sectional tube of the torus.
(E) At any point on the surface two principle curvatures $C_1$ and $C_2$ parametrize the shape that can be combined into the mean curvature $H$ and Gaussian curvature $K$.
(F) Normalized squared mean curvature on the Clifford torus, which has a toroidal aspect ratio of $\rho=\Rring/\rcross=\sqrt{2}$.
(G) Normalized Gaussian curvature on the Clifford torus.
(H) Toroidal vesicles for pure bolalipid membranes at $\km=\unit[0]{}\kBT$ (top), at
$\km=\unit[1]{}\kBT$ (center) and a mixture membrane with bilayer content of $f_\mathrm{bi}=0.7$ and bolalipids at $\km=\unit[2]{}\kBT$. 
\label{fig:Figure1}}
\end{figure*}

\section{Introduction}
Cells are defined by their boundary -- the cell membrane \cite{alberts2015}.  
This fluid membrane is composed of lipid molecules that, across the tree of life, self-assemble into distinct geometric structures  \cite{gould2018}. 
In bacteria and eukaryotes, membranes are typically organized as bilayers formed from single-headed bilayer lipids, which contain one hydrophilic head group attached to two hydrophobic tails \cite{albers2011}. 
By contrast, in archaea -- the third domain of life -- both bilayer and monolayer geometric structures are observed.
This structural diversity arises because archaeal membranes are composed of mixtures of bilayer lipids and double-headed bolalipids, which possess two hydrophilic head groups connected by two hydrophobic tails \cite{albers2011}.
Owing to their molecular geometry, bolalipids are believed to adopt two distinct conformations within membranes: a straight (trans-membrane) conformation in which both hydrophilic head groups occupy opposite membrane leaflets and a u-shaped (looped) conformation in which both head groups occupy the same membrane leaflet \cite{brownholland2009,bhattacharya2024}.
This conformational versatility is thought to enable archaeal membranes to interpolate between bilayer and monolayer membrane structures, governed by the bilayer lipid content and the abundance of bilayer-mimicking u-shaped bolalipids.   

Archaeal membranes, like those of bacteria and eukaryotes, continuously undergo remodeling processes and topological transformations such as fusion and fission events \cite{tarrason2020,liu2021}.
The mechanisms available for these transformations depend on whether the membrane adopts a bilayer or monolayer structure, which governs the accessibility of processes such as curvature generation, leaflet area asymmetry and hemifusion \cite{warner2012,spencer2024}.
To theoretically investigate such processes, fine-grained models are required \cite{munoz2025}, as continuum theories often do not capture internal membrane structure \cite{seifert1997,deserno2015}.
While curvature generation and membrane remodeling have been extensively studied in bilayer membranes \cite{mcmahon2005,zimmerberg2006,frey2021}, the physical principles governing archaeal membranes remain largely unexplored.
In particular, it is unclear how archaeal membranes with bilayer or monolayer geometric structure remodel and undergo topological transitions, or how membrane structure couples to externally imposed curvature. 

Here, we investigate how the structure of archaeal membranes, together with external curvature, governs shape stability, lipid organization and topological remodeling.
Analogous to living archaea, which can express bolalipids with different rigidities \cite{chong2010} and regulate the fraction of bilayer lipids in their membranes \cite{tourte2020}, our model allows independent control over lipid stiffness and bilayer fraction. 
We conduct coarse-grained molecular dynamics simulations of toroidal vesicles, which represent energetically metastable membrane configurations.
The geometry enables us to identify the conditions under which bilayer and monolayer membranes undergo shape transitions.  
Moreover, these toroidal vesicles, owing to their small size ($\sim\unit[70]{nm}$ diameter), are highly curved, and encompass all relevant combinations of mean and Gaussian curvatures, enabling a systematic investigation of how curvature influences lipid organization.

We conclude that both membrane structure and the membrane curvature jointly control shape stability and lipid organization during membrane remodeling.
Specifically, our simulations show that toroidal vesicles of soft bilayer membranes, are stable regardless of whether they are composed of mixtures of straight and u-shaped bolalipids or of bilayer lipids.
In these mixture membranes, both u-shaped bolalipids and bilayer lipids accumulate in regions of high mean curvature, independent of Gaussian curvature. 
By contrast, toroidal vesicles are unstable in rigid monolayer membranes composed of bolalipids that preferentially adopt a straight conformation, as well as in mixture membranes with low bilayer lipid content.
In these cases, toroidal vesicles undergo shape transitions and form either spherical vesicles or flat membrane sheets.
During such shape transitions, membrane pores can emerge, controlled by the bolalipid rigidity and the bilayer fraction. 
Notably, pores form heterogeneously on the inner and outer surfaces of the torus, where mean curvature is maximal, suggesting that pore formation serves to relieve curvature-induced stress.

\section{Modeling archaeal membrane vesicles of toroidal geometry}
Archaeal membranes can be composed of single-headed bilayer lipids and double-headed bolalipids, and can adopt either a bilayer structure similar to bacterial and eukaryotic membranes or a monolayer structure (Fig.~\ref{fig:Figure1}A). 
To model archaeal membranes, we extend the bilayer lipid model by Cooke and Deserno \cite{cooke2005} and build upon our previous work \cite{amaral2023}.
In the model, bilayer lipids are represented by 3 connected, volume-excluded beads.
The molecular rigidity is controlled by a single harmonic angle potential between the head bead and the terminal tail bead, parametrized by $\ko$ (Fig.~\ref{fig:Figure1}B, left).
Bolalipids are modeled analogously as 6 connected, volume-excluded beads, with two harmonic angle potentials between the two head beads and the corresponding tails beads, also parametrized by $\ko$  (Fig.~\ref{fig:Figure1}B, right). 
To account for the increased rigidity associated with cyclopentane rings found in archaeal bolalipids \cite{chong2010}, we further introduce two additional harmonic angle potentials between pairs of tail beads. 
This interaction is parametrized by $\km$ and allows systematic tuning of bolalipid stiffness.

Lipid tail beads interact via an effective attraction due to the hydrophobic effect of the implicit solvent. 
This attraction is characterized by the parameters $\omega$, which sets the interaction range and $\epsilon_\mathrm{p}$, which determines the interaction strength (Fig.~\ref{fig:Figure1}B).
Tail beads of both bilayer lipids (cyan) and bolalipids (pink) attract one another, whereas head beads (gray) interact solely through volume exclusion.
A complete description of the model and interaction potentials is provided in the supplementary material.

A bolalipid can mainly adopt two conformations within a membrane, which we quantify by the angle $\theta$ between the two head beads (Fig.~\ref{fig:Figure1}C).
In the u-shaped (looped) conformation, both heads beads occupy the same membrane leaflet, whereas in the straight conformation the head beads occupy opposing leaflets.

To investigate the effect of membrane curvature, we assemble lipids into toroidal vesicles.
The torus surface can be parametrized in Cartesian coordinates as:
\begin{align}
x&=\left ( \Rring+\rcross \sin \vartheta \right ) \cos \varphi 
\nonumber
\\
y&=\left ( \Rring+\rcross \sin \vartheta \right ) \sin \varphi
\nonumber
\\
z&=\rcross \cos \vartheta \, .
\end{align}
Here, $\Rring$ and $\rcross$ are the ring (major) radius and cross-section (minor) radius, respectively and $\vartheta \in [-\pi,\pi]$ and $\varphi \in [0,2\pi]$ are the toroidal (polar-like) angle around the cross-sectional tube and the azimuthal angle around the torus' major ring (Fig.~\ref{fig:Figure1}D).
The surface parametrization implies that one radius -- here, the cross-section radius $\rcross$ -- sets the length scale of the torus, while the dimensionless toroidal aspect ratio (the ratio of radii), $\rho=\Rring/\rcross$, determines its overall shape.
To investigate the effect of membrane curvature and to limit system size, we select a relatively modest cross-section radius, $\rcross=\unit[15]{\sigma}$, which enforces a high membrane curvature; here, $\sigma$ is the size of a lipid bead and the simulation unit of length, approximately corresponding to $\sim \unit[1]{nm}$.
For the torus shape, we adopt the Clifford torus with the toroidal aspect ratio $\rho=\sqrt{2}$, which minimizes the bending energy of the membrane according to the Helfrich Hamiltonian \cite{ou-yang1990}.
With these parameters, the resulting toroidal vesicle has a diameter of roughly $2(\Rring+\rcross)\approx\unit[72]{nm}$.

\begin{table}[b]
\caption{Model parameter values.}
\label{table:Parameters}
\begin{tabular}{cc}
\hline
Parameter & Value  \\
\hline
Simulation unit of energy & 
$\unit[\kBT]{}$ \\
Simulation unit of length & 
$\unit[\sigma]{}$ \\
Simulation unit of time 
& 
$\unit[t_0]{}$ \\
Bolalipid rigidity $\km$ & 
$\unit[0-5]{}\kBT$ \\
Interaction strength $\epsilon_\mathrm{p}\;\;\;\;$ & 
$\kBT/(1.3+\km/(\unit[10]{}\kBT))$ \\
Lipid bead size & 
$\unit[\sigma]{}$ \\
Interaction range $\omega$ & 
$\unit[1.5]{\sigma}$ \\
Timestep $t_\mathrm{s}$ & 
$\unit[0.01]{t_0}$ \\
Simulation time $\tau$
& 
$\unit[4 \times 10^4]{t_0}$ \\
Simulation beads & 
$\unit[54000]{}$ \\
\hline
\end{tabular}
\end{table}

Comparing the bending and topological energies for the shape transition from a Clifford torus to a sphere, in the absence of a negative preferred membrane curvature, we obtain the fission energy \cite{ou-yang1990}:
\begin{equation}
E_\mathrm{fission}^{\circ \rightarrow \bullet}=4 \pi \kappa  \left(2-\pi +\frac{\bar{\kappa}}{\kappa} \right)\, .
\label{eq:fission_energy_gain}
\end{equation}
Here, $\kappa$ is the bending rigidity of the membrane and $\bar{\kappa}$ is the Gaussian curvature modulus.
Using the Cooke and Deserno model for bilayer membranes \cite{cooke2005} and its extension to bolalipid membranes \cite{amaral2023}, the ratio of the Gaussian curvature modulus to the bending rigidity was found to be $\bar{\kappa}/\kappa \sim -1$ for bilayers \cite{hu2012} and $\bar{\kappa}/\kappa \sim -0.5$ for flexible bolalipid membranes at vanishing molecular rigidity ($\km=\unit[0]{}\kBT$) \cite{amaral2023}. 
The values fall within the range $-0.3 \le \kappa/\kappa  \le -1$, commonly reported for lipid membranes \cite{hu2012,deserno2015}.
Although intermediate energy barriers may hinder the transition, Eq.~(\ref{eq:fission_energy_gain}) predicts an energy gain when the membrane transitions from a torus to a sphere.
Eq.~(\ref{eq:fission_energy_gain}) also suggests that the toroidal vesicle can be destabilized by incorporating bolalipids of higher molecular rigidity $\km$ or by decreasing the bilayer fraction $\fbi$ in a mixture membrane, which both increase the bending rigidity $\kappa$ \cite{amaral2023}.
We thus expect the Clifford torus to be energetically metastable, with the fission transition to a spherical vesicle occurring depending on both the membrane rigidity and the archaeal membrane structure.

The torus exhibits spatially varying curvatures along its surface, with the mean curvature $H$ and Gaussian curvature $K$ given by \cite{wohlert2006}:
\begin{align}
H&=\frac{1}{2}\left(C_1 +C_2\right)=\frac{\Rring+2\rcross \sin \vartheta}{2 \rcross 
\left ( \Rring+\rcross \sin \vartheta \right )}    
\\
K&=C_1 C_2=\frac{\sin \vartheta}{\rcross (\Rring+\rcross \sin \vartheta)} \, ,
\label{eq:HandK}
\end{align}
where $C_1$ and $C_2$ are the principal curvatures along the torus surface(Fig.~\ref{fig:Figure1}E). 
We characterize the toroidal vesicle geometry using the squared mean curvature $H^2$ and the Gaussian curvature $K$. 
While the squared mean curvature is always non-negative, the Gaussian curvature can be positive, negative or zero.
Apart from the trivial case of a flat membrane with vanishing squared mean curvature and Gaussian curvature ($H^2=0$ and $K=0$), all other combinations of curvatures manifest themselves on the Clifford torus.
Specifically, we observe $H^2>0$ and $K>0$ at the outer surface (sphere-like), $H^2>0$ and $K=0$ at the top and bottom (cylinder-like), $H^2>0$ and $K<0$ at the inner surface (saddle-like) and $H^2=0$ and $K<0$ at the interface between inner surface and top/bottom (catenoid-like). These regions are illustrated in Fig.~\ref{fig:Figure1}F and G.

We simulate toroidal vesicles composed of pure bolalipids with varying molecular rigidity $\km$ (see Fig.~\ref{fig:Figure1}H, top and center), as well as vesicles containing mixtures with varying bilayer fraction $\fbi$ and relatively stiff bolalipids with $\km=\unit[2]{}\kBT$ (see Fig.~\ref{fig:Figure1}H, bottom).
Simulation parameters for both pure and mixed membranes are listed in Table~\ref{table:Parameters}.
The dynamics of lipids in our coarse-grained molecular dynamics simulation are obtained via Langevin dynamics (see supplementary material).

\begin{figure*}
\centering
\includegraphics[width=.95\textwidth]{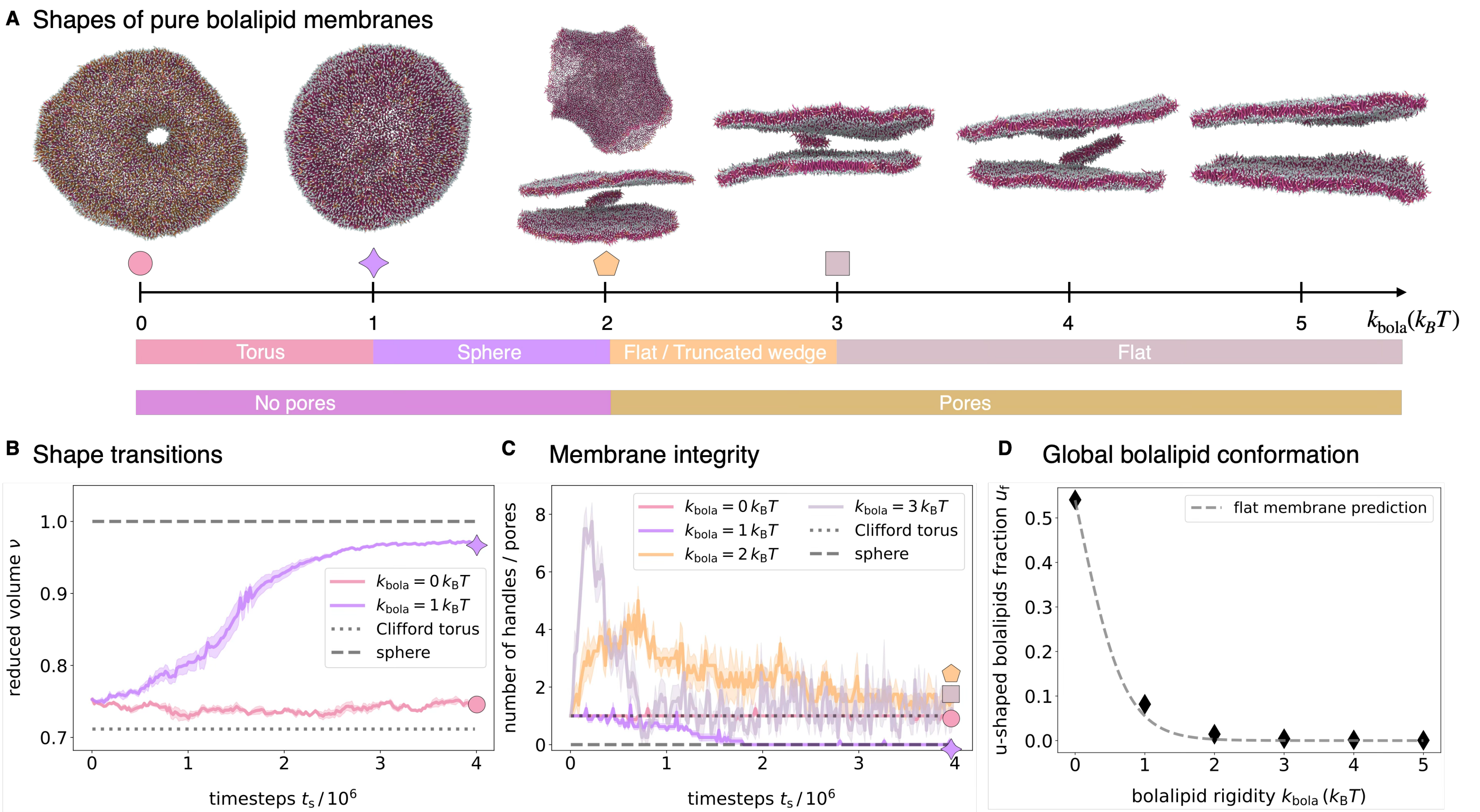}
\caption{Bolalipid rigidity controls stability of toroidal vesicles composed of bolalipids.
(A) Shape diagram of pure bolalipid membranes as a function of the bolalipid rigidity $\km$. 
For $\km=\unit[0]{}\kBT$, we observe toroidal vesicle shapes. 
With increasing values of $\km$, we observe spherical vesicles ($\km=\unit[1]{}\kBT$), flat membrane sheets or truncated wedges (ratio 2:2 for $\km=\unit[2]{}\kBT$) and flat membrane sheets ($\km > \unit[2]{}\kBT$).
Membrane pores can only be observed for $\km\ge \unit[2]{}\kBT$.
(B) Reduced volume $\nu$ along the simulation trajectory for different values of the bolalipid rigidity $\km$.
(C) Number of handles or pores in the membrane along the simulation trajectory for different values of the bolalipid rigidity $\km$.
(D) Global u-shaped bolalipid fraction $\uf$ as a function of the bolalipid rigidity $\km$. 
Dashed gray line shows $\uf(\km)$, fitted for a flat membrane. 
(B-D) Each data point shows an average over $N_\mathrm{seeds}=4$ random initial seeds.  
\label{fig:Figure2}}
\end{figure*}

\section{Bolalipid rigidity determines stability of toroidal vesicles of pure bolalipid membranes}
We first assess the stability of Clifford toroidal vesicles composed of pure bolalipid membranes.
The vesicle shape at the end of the simulated trajectories depends strongly on the bolalipid rigidity $\km$ (Fig.~\ref{fig:Figure2}A, Video S1). 
For vanishing bolalipid rigidity ($\km=\unit[0]{}\kBT$), the toroidal vesicles remain stable. 
At moderate bolalipid rigidity ($\km=\unit[1]{}\kBT$), we observe a transition to a spherical vesicle geometry. 
The transition occurs via membrane fission at the center of the torus, causing the stalk of the torus to vanish.
Upon further increasing the bolalipid rigidity ($\km\ge\unit[2]{}\kBT$), the membrane becomes unstable and ruptures: pores form on the outer surface of the toroidal vesicles and the membrane ultimately flattens. 
These observations can be rationalized within continuum membrane theory. 
Increasing the bolalipid rigidity $\km$ leads to higher bending rigidity $\kappa$ \cite{amaral2023}, which enhances the energy gain associated with membrane fission (see Eq.~(\ref{eq:fission_energy_gain})).
At sufficiently high bolalipid rigidity $\km$, however, membrane bending becomes energetically prohibitive, resulting in membrane rupture rather than a smooth topological transition.

To quantify the smooth shape transitions of the vesicles along the simulated trajectories, we use the dimensionless reduced volume 
\begin{equation}
\nu=\frac{3 V}{4\pi  \left(\frac{A}{4\pi}\right)^{3/2}} \, ,
\end{equation}
which measures the ratio of the enclosed volume $V$ to the surface area $A$.
The reduced volume is one for a sphere ($\nu=1$) and smaller than one for all other closed shapes ($\nu<1$).
We compute the reduced volume $\nu$ by mapping the membrane surface onto a triangulated mesh and evaluating the enclosed volume $V$ and the surface area $A$ at every tenth timestep of the trajectory (see supplementary material).
Owing to this procedure, the reduced volume $\nu$ can be determined only if the membrane is devoid of pores, restricting the analysis to vanishing and moderate bolalipid rigidities ($\km\le\unit[1]{}\kBT$).

The time evolution of the reduced volume $\nu$, averaged over independent simulations seeds, is shown in Fig.~\ref{fig:Figure2}B for different values of the bolalipid rigidity $\km$.
For vanishing bolalipid rigidity ($\km=\unit[0]{}\kBT$, solid red line), the reduced volume $\nu$ remains approximately constant throughout the trajectory. The small deviation from the theoretical value for a Clifford torus, $\nu_\mathrm{Clifford}\approx 0.712$ (dotted gray line), is expected because the membrane has a finite thickness (see supplementary material). 
For a moderate bolalipid rigidity ($\km=\unit[1]{}\kBT$, solid purple line), the reduced volume starts close to the theoretical value $\nu_\mathrm{Clifford}$ and gradually approaches 1 (dashed gray line), indicating a transition from a Clifford torus to a spherical vesicle within simulation time.

To quantify how vesicle shape transitions proceed, we monitor the number of handles or pores in the membrane throughout the trajectory. 
To this end, we employ Euler's polyhedron formula, which relates the genus $g$ (the number of handles or pores) of $N$ closed surfaces to its mesh topology via the equation  $2(N-g)=V-E+F$, where $V$, $E$ and $F$ are the number of vertices, edges and faces obtained by remeshing the vesicle surface.
As long as the reduced volume $\nu$ can be determined, we reconstruct two separate surface meshes corresponding to the outer and inner membrane leaflets.
To evaluate the number of handles, we  account for both leaflets.
Once the membrane ruptures and the reduced volume $\nu$ can no longer be evaluated, the two meshes merge into a single surface $N=1$.
In this regime, the genus $g$ directly represents the number of pores that perforate the membrane (see supplementary material).

The time evolution of the number of handles or pores for different values of the bolalipid rigidity $\km$, averaged over independent simulations seeds, is shown in Fig.~\ref{fig:Figure2}C. 
For vanishing bolalipid rigidity ($\km=\unit[0]{}\kBT$, solid red line), one handle persists throughout the entire trajectory, corresponding to the handle of the Clifford torus (dotted gray line).
For moderate bolalipid rigidity ($\km=\unit[1]{}\kBT$, solid purple line), the number of handles decreases from one to zero as the toroidal vesicle transitions to a spherical vesicle (dashed gray line).
In contrast, for stiff bolalipids ($\km\ge\unit[2]{}\kBT$, yellow and gray solid lines), the number of pores initially increases and then rapidly decreases, indicating membrane rupture.

By comparing the vesicle shapes shown in Fig.~\ref{fig:Figure2}A, we observe that u-shaped bolalipids (orange) become less prevalent as the bolalipid rigidity $\km$ increases.
In contrast, the fraction of straight bolalipids (pink) increases as we observe shape transitions.
To quantify the effect, we measure the global fraction of u-shaped bolalipids $\uf$ in the steady state, as well as the membrane relaxation timescale, as a function of the bolalipid rigidity $\km$ (Fig.~S1).
The steady state results of the u-shaped bolalipid fraction $\uf$, averaged over independent simulations seeds, are shown in Fig.~\ref{fig:Figure2}D. 
We find that $\uf$ decreases monotonically with increasing bolalipid rigidity $\km$, in agreement with previous measurements for flat membranes (gray line, see supplementary material) \cite{amaral2023}. 
However, for toroidal and spherical vesicles ($\km=\unit[0]{}\kBT$ and $\km=\unit[1]{}\kBT$), the u-shaped bolalipid fraction $\uf$ is slightly enhanced compared to the prediction for the flat membrane.
Moreover, during the transition from a toroidal vesicle to a spherical vesicle at moderate bolalipid rigidity ($\km=\unit[1]{}\kBT$), the fraction of u-shaped bolalipids decreases over time (Fig.~S1B).
This observation shows that membrane shape and, more generally, membrane curvature globally promote the stabilization of u-shaped bolalipid conformations.

\begin{figure*}
\centering
\includegraphics[width=.95\textwidth]{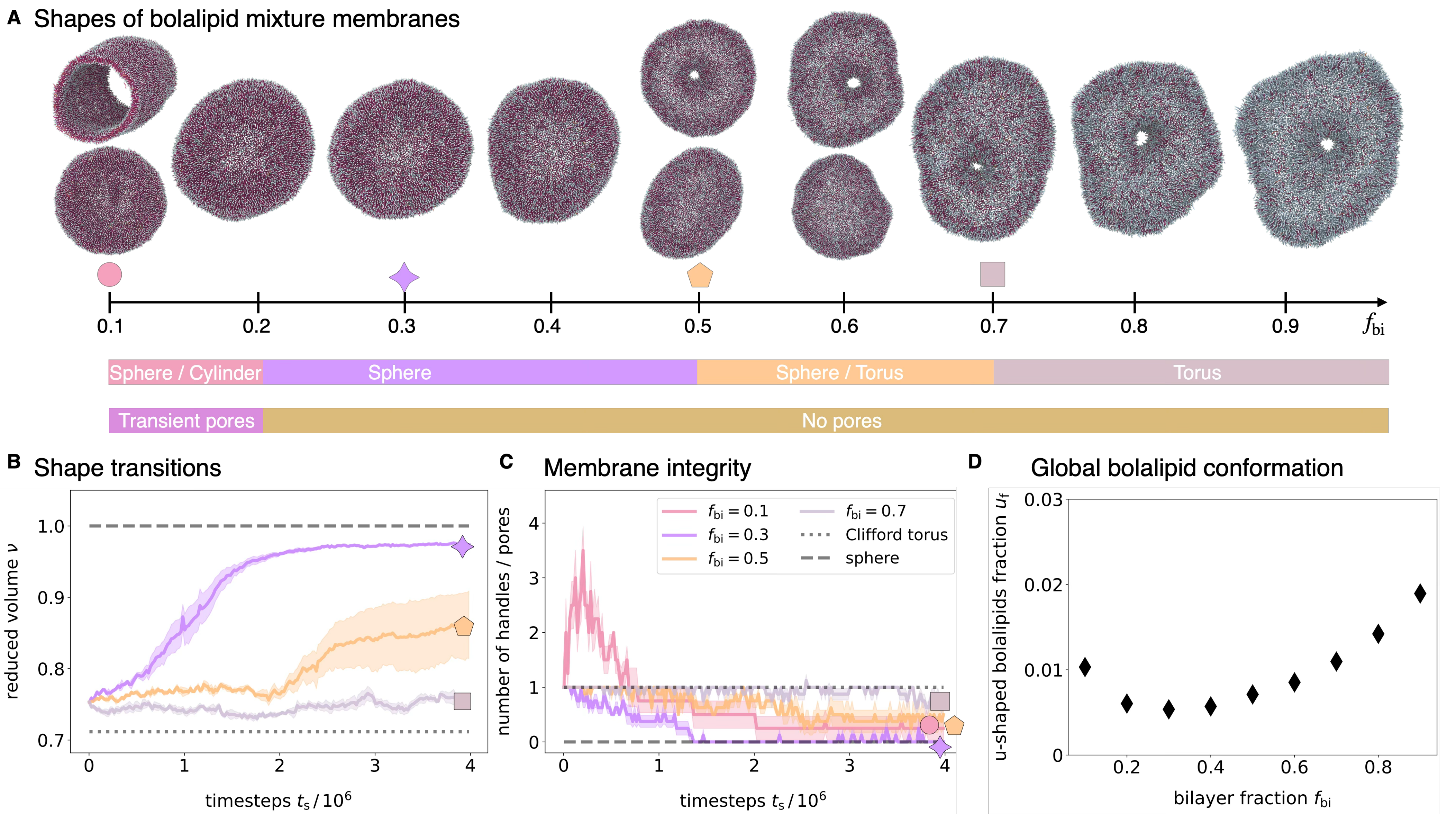}
\caption{Bilayer lipid fraction controls stability of toroidal vesicles of archaeal mixture membranes composed of bilayer and bolalipids.
(A) Shape diagram of mixture membranes as a function of bilayer fraction $\fbi$. 
For $\fbi=0.1$, we observe spherical or cylindrical membrane shapes (ratio 3:1). 
With increasing values of $\fbi$, we first observe spherical membrane shapes, followed by vesicles of torus shape.
For $\fbi=0.5$ and $\fbi=0.6$, we observe either spherical or toroidal vesicles (ratio 2:2) as final state in independent simulation runs.
For $\fbi=0.1$, the transition from the initial toroidal membrane shape is accompanied by the appearance of transient membrane pores, which we do not observe otherwise.
(B) Reduced volume $\nu$ along the simulation trajectory for different bilayer fractions $\fbi$.
(C) Number of handles or pores in the membrane along the simulation trajectory for different bilayer fractions $\fbi$.
(D) Global u-shaped bolalipid fraction $\uf$ as a function of $\fbi$ for mixture membranes of bilayer and bolalipids.
(B-D) Each data point shows an average over $N_\mathrm{seeds}=4$ random initial seeds.  
\label{fig:Figure3}}
\end{figure*}

\section{Bilayer lipid fraction determines stability of toroidal vesicles of archaeal mixture membranes}
Archaeal membranes are typically composed of both bilayer and bolalipids \cite{tourte2020}.
To investigate their shape stability, we next consider toroidal vesicles formed from mixed archaeal membranes consisting of relatively stiff bolalipids ($\km=\unit[2]{}\kBT$) and a variable bilayer lipid fraction $\fbi$. 
We first construct a state diagram based on the vesicles shapes observed at the end of the simulated trajectories (Fig.~\ref{fig:Figure3}A, Video S2).
At low bilayer lipid fraction ($\fbi=0.1$), we observe shape transitions of the toroidal vesicle accompanied by membrane pore formation. 
Depending on the pore location, the final state is either a spherical vesicle (pores form on the outer surface) or a cylinder vesicle (pores form at opposite ends).
For intermediate bilayer fractions $\fbi=0.2-0.4$, no membrane pores are observed; instead, the vesicle shape evolves smoothly into a sphere. 
At bilayer fractions of $\fbi=0.5$ and $\fbi=0.6$, independent simulation runs yield either toroidal or spherical vesicles as final states.
For large bilayer fractions ($\fbi\ge0.7$), only stable toroidal vesicles are observed. 

To quantify the vesicle shape, we evaluate the reduced volume $\nu$, averaged over multiple simulation trajectories, for a varying  bilayer fraction $\fbi$. 
The results are shown in Fig.~\ref{fig:Figure3}B.
For a small bilayer fraction ($\fbi=0.3$, solid purple line), the reduced volume $\nu$ increases from the theoretical value of the Clifford-torus $\nu_\mathrm{Clifford}$ (dotted gray line) to $\nu=1$, confirming a transition to a spherical vesicle (dashed gray line).
The reduced volume can be computed along the entire trajectory (the line is continuous), indicating that the membrane remains pore-free throughout the simulation trajectory.
For an intermediate bilayer fraction ($\fbi=0.5$, yellow solid line), the reduced volume $\nu$ fluctuates considerably, reflecting the occurrence of either toroidal or spherical vesicles across independent simulation runs.
Finally, at high bilayer fraction ($\fbi=0.7$, gray solid line), the reduced volume $\nu$ remains constant and close to the theoretical value of the Clifford torus $\nu_\mathrm{Clifford}$, demonstrating that the toroidal vesicle is stable and that a transition to a spherical shape no longer occurs.

To assess membrane integrity, we monitor the time evolution of the number of handles or pores in the vesicle membranes, averaged over independent simulations seeds, throughout the simulation trajectories  (Fig.~\ref{fig:Figure3}C). 
For a low bilayer fraction ($\fbi=0.1$, red solid line), the genus $g$ initially increases and subsequently decreases to zero. 
This behavior confirms that the toroidal vesicle transforms into a spherical vesicle (gray dashed line), with the transformation proceeding via the transient opening of membrane pores.
For intermediate bilayer fractions $\fbi=0.3-0.5$ (purple and yellow solid lines), the number of handles decreases along the trajectory, indicating a smooth topological transition from a toroidal to a spherical vesicle without pore formation.
In contrast, at a high bilayer fraction ($\fbi=0.7$, gray solid line), the number of handles remains approximately one, confirming that the toroidal vesicle is stable and persists throughout the simulation (gray dotted line). 

To analyze membrane organization, we examine the fraction of u-shaped bolalipids and the distribution of bilayer lipids in Fig~\ref{fig:Figure3}A. 
To quantify the abundance of u-shaped bolalipids, we analyze trajectories in steady state and determine the global fraction of u-shaped bolalipids $\uf$ as well as the relaxation timescale, as functions of the fraction of bilayer lipids $\fbi$ (Fig.~S2).
The steady state values of the u-shaped bolalipid fraction $\uf$, averaged over multiple simulations seeds, as a function of the bilayer fraction $\fbi$ are shown in Fig.~\ref{fig:Figure3}D. 
Overall, the fraction of u-shaped bolalipids $\uf$ remains rather small ($\le 2\%$).
Notably, however, the fraction of u-shaped bolalipids $\uf$ exhibits a non-monotonic dependence on the bilayer fraction $\fbi$.
For low bilayer fractions ($\fbi=0.1-0.4$), the u-shaped bolalipid fraction $\uf$ decreases as bilayer lipids increasingly replace the structural role of  u-shaped bolalipids.
In contrast, at higher bilayer fractions ($\fbi=0.5-0.9$), the toroidal vesicle becomes progressively more stable and the torus geometry promotes the formation of u-shaped bolalipids.

\begin{figure*}
\centering
\includegraphics[width=.95\textwidth]{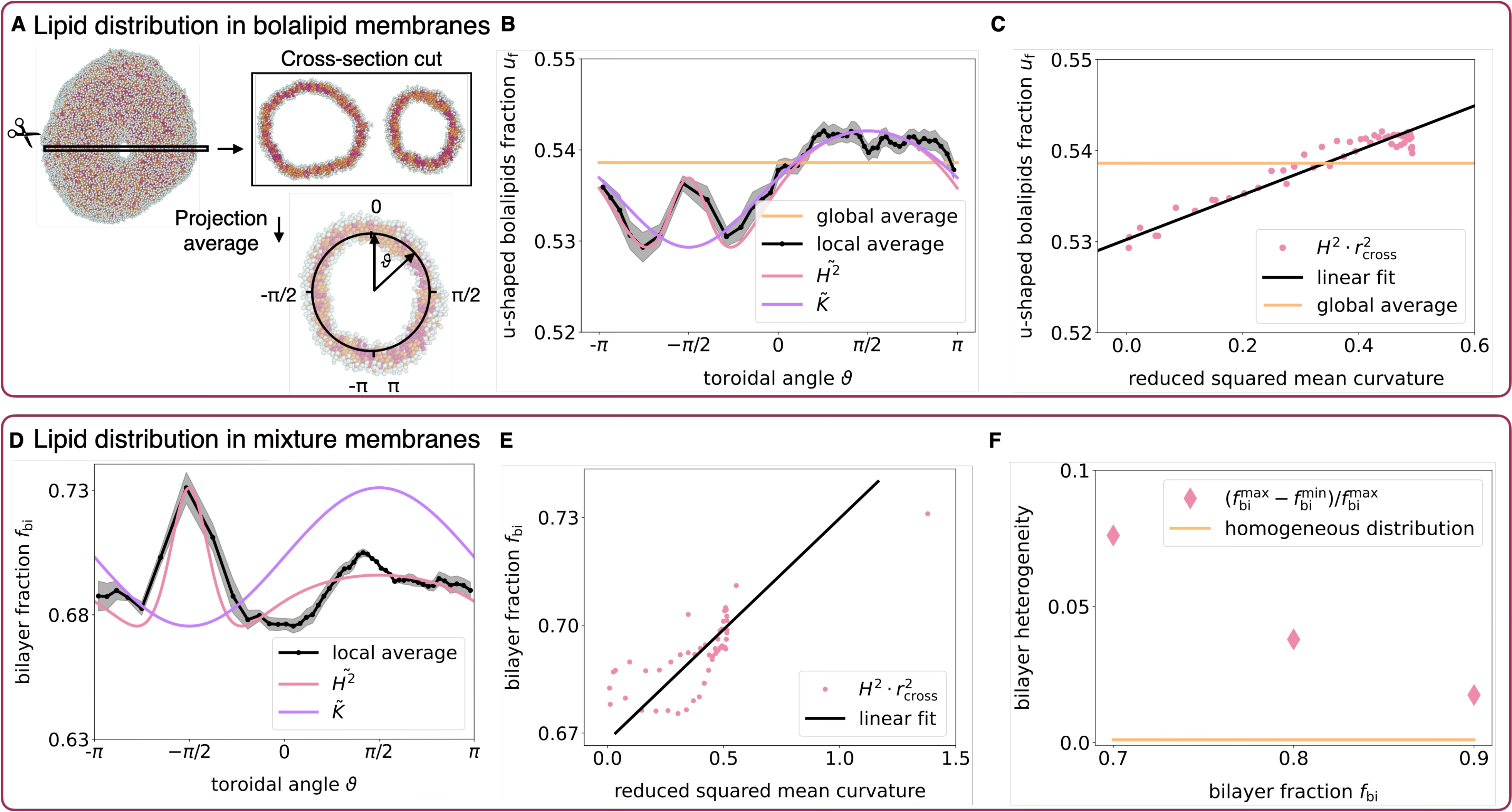}
\caption{Curvature controls lipid distribution in pure bolalipid and archaeal mixture membranes.
(A) For $\km=\unit[0]{}\kBT$, we observe membrane vesicles of torus shape (left). 
For the torus shape, the u-shaped bolalipid fraction $\uf$ is mapped on the cross-section circle (right), averaged  and represented by the toroidal angle $\vartheta$ (bottom).
(B) u-shaped bolalipid fraction $\uf$ as a function of $\vartheta$ (solid black line) with standard error (gray area).
Along with $\uf$, the scaled squared mean curvature $\tilde{H^2}$ (solid red line) and the scaled Gaussian curvature $\tilde{K}$ (solid purple line) are plotted.
The global value along the torus is shown by the solid orange line.
(C) The u-shaped bolalipid fraction $\uf$ as a function of the reduced squared mean curvature $H^2 \cdot \rcross^2$ (red) and a linear fit to the data (solid black line).
(B and C) Each data point shows an average over $N_\mathrm{seeds}=16$ random initial seeds.  
(D) For a bilayer fraction of $\fbi=0.7$, we observe membrane vesicles of torus shape. 
The averaged bilayer fraction $\fbi$ as a function of the toroidal angle $\vartheta$ (solid black line) with standard error (gray area).
Along with $\fbi$, the scaled squared mean curvature $\tilde{H^2}$ (solid red line) and the scaled Gaussian curvature $\tilde{K}$ (solid purple line) are plotted.
(E) The bilayer fraction $\fbi$ as a function of the reduced squared mean curvature $H^2 \cdot \rcross^2$ (red) and a linear fit to the data (solid black line).
(F) Bilayer heterogeneity (for definition see text) as a function of the bilayer lipid fraction $\fbi$.
A completely homogeneous membrane would show a bilayer heterogeneity of 0 (solid orange line).
(D-E) Each data point shows an average over $N_\mathrm{seeds}=4$ random initial seeds.  
\label{fig:Figure4}}
\end{figure*}

\section{Curvature controls lipid sorting in archaeal membranes}
To investigate the spatial distribution of lipids in pure bolalipid membranes, we consider stable toroidal vesicles at vanishing bolalipid rigidity ($\km=\unit[0]{}\kBT$). 
We first verify that the toroidal vesicles remain stable along the simulation trajectories.
For this purpose, we monitor the reduced volume $\nu$ (Fig.~\ref{fig:Figure2}B), the global amount of u-shaped bolalipids $\uf$ (Fig.~S1A), the ring radius $\Rring$ (Fig.~S3A), the cross section radius $\rcross$ (Fig.~S3B), the toroidal aspect ratio $\rho$ (Fig.~S3C) and the toroidal hole radius $\Rring - \rcross$ (Fig.~S3D). 
Since all these quantities exhibit only minor variations over time, we consider the torus shape to be stable.
This allows us to average over the trajectories and quantify the spatial fraction of u-shaped bolalipids as a function of position on the vesicle surface.

Given that the toroidal vesicle shape is stable, we determine the spatial distribution of u-shaped bolalipids along the membrane surface (see supplementary material). 
To this end, we average over $N_\mathrm{seeds}=16$ independent trajectories of Clifford tori in steady state (Fig.~\ref{fig:Figure4}A left), project the bolalipid coordinates onto a cross-sectional circle (Fig.~\ref{fig:Figure4}A right) and represent each position by the toroidal (polar-like) angle $\vartheta$ (Fig.~\ref{fig:Figure4}A bottom).
Fig.~\ref{fig:Figure4}B shows that the spatial u-shaped bolalipid fraction $\uf$ is heterogeneously distributed along the toroidal angle $\vartheta$ (black line with standard error shown as the shaded gray area) and deviates considerably from the global average of the u-shaped bolalipid fraction $\uf$ (orange line).

To test whether the mean or Gaussian curvature are correlated with the local u-shaped bolalipid fraction $\uf$, we fit the scaled squared mean curvature $\tilde{H^2}$ (red line) and scaled Gaussian curvature $\tilde{K}$ (purple line) of the Clifford torus, defined as 
\begin{align}
\tilde{H^2} (\vartheta,\rho)&=\min (\uf)+(\max (\uf)-\min (\uf)) \frac{H^2}{ \max (H^2)}
\label{eq:scaled_H}
\\
\tilde{K}(\vartheta,\rho)&= \min (\uf)+(\max (\uf)-\min (\uf)) \times 
\nonumber
\\
&\times \frac{K-\min (K)}{ \max (K)-\min (K)}
\, ,
\label{eq:scaled_K}
\end{align}
where $H$ and $K$ are the mean and Gaussian curvature on a torus (see Eq.~(\ref{eq:HandK})) and $\max(\cdot)$ and $\min(\cdot)$ denote the domain maximum and minimum values.
This scaling ensures that $\tilde{H^2}$ and $\tilde{K}$ are bounded between $\min (\uf)$ and $\max (\uf)$. 
Since the shape of the torus can vary slightly over time (Fig.~S3), we treat the toroidal aspect ratio $\rho$ as a free fit parameter in Eqs.~(\ref{eq:scaled_H}) and (\ref{eq:scaled_K}).
Fig.~\ref{fig:Figure4}B shows that the u-shaped bolalipid fraction $\uf$ scales closely with the squared mean curvature $\tilde{H^2}$, whereas no correlation is observed with the Gaussian curvature $\tilde{K}$.
This indicates that the local squared mean curvature $\tilde{H^2}$ is the primary geometric factor controlling the u-shaped bolalipid fraction $\uf$.
From the fit, we obtain a toroidal aspect ratio of $\rho=1.48$, which agrees with the value of $\rho=1.39$ obtained from a direct fit of the torus geometry  (Fig.~S3C).
We conclude that membrane geometry, specifically the local squared mean curvature, governs lipid sorting in these vesicles.  

To rationalize the observed lipid sorting in terms of squared mean curvature, we turn to continuum modeling, following previous theoretical studies on lipid sorting in bilayer membranes \cite{cooke2006,derganc2007,callan-jones2011}.
We consider the energetic cost of bending a membrane patch of area $A$ and constant mean curvature $H$
\begin{equation}
E=2 \kappa H^2 A\, ,
\end{equation}
where $\kappa$ is the bending rigidity of the membrane.
We assume that $\kappa$ depends on the local u-shaped bolalipid fraction $\uf$ in curved membrane patches.
Expanding the bending rigidity to linear order around its value in a flat reference state, we write $\kappa \approx\kappa^0 + \kappa^\mathrm{c} (\uf^0-\uf)$, where $\uf^0$ is the bolalipid fraction in the flat membrane, $\kappa^0$ and $\kappa^\mathrm{c}$ are the bending rigidities in the flat and curved states, respectively.
Thus, membrane patches exhibit a reduced bending rigidity $\kappa$ with increased u-shaped bolalipid fraction $\uf$, whereas patches become stiffer when the u-shaped bolalipid fraction $\uf$ is decreased.

\begin{table}[b]
\caption{Toroidal aspect ratio $\rho=\Rring/\rcross$ derived from the analysis of the torus geometry (Figs.~S3 and S4-6) and the analysis of $\uf$ (Fig.~\ref{fig:Figure4}B) and $\fbi$ (Fig.~\ref{fig:Figure4}D and Fig.~S7).}
\label{table:rho}
\begin{tabular}{ccc}
\hline
Simulation & $\rho$ (geometry) &  $\rho$ (lipid composition)  \\
\hline
Bolalipid torus & $\rho=1.39$ & $\rho=1.48$\\
Bilayer torus $\fbi=0.7$ & $\rho=1.36$ & $\rho=1.30$\\
Bilayer torus $\fbi=0.8$ & $\rho=1.38$ & $\rho=1.20$\\
Bilayer torus $\fbi=0.9$ & $\rho=1.36$ & $\rho=1.31$\\
\hline
\end{tabular}
\end{table} 

Furthermore, we assume that straight and u-shaped bolalipids behave as an ideal lattice gas within the membrane plane. 
In this case, the mixing entropy, in the absence of molecular interactions, is given by
\begin{equation}
S=- k_\mathrm{B} \left ( (N-N_\mathrm{u}) \ln (1-\uf) + N_\mathrm{u} \ln \uf \right) \, ,
\end{equation}
with $N$ and $N_\mathrm{u}$ being the number of bolalipids in total and the number of u-shaped bolalipids, respectively.
We assume that the mixing entropy is maximized in the homogeneous flat state at $\uf^0$ and we expand $S$ around this state:
\begin{equation}
-TS \approx-T S^0 + \frac{1}{2}k^\mathrm{s} (\uf^0-\uf)^2 \, ,
\end{equation}
where $S^0$ is the entropy of the homogeneous state and $k^\mathrm{s}$ is an expansion coefficient.
Minimizing the free energy $F=E-TS$ with respect to the u-shaped bolalipid fraction $\uf$ then yields
\begin{equation}
\uf=\uf^0+\zeta H^2 \, ,
\label{eq:uf_with_H2}
\end{equation}
with $\zeta=2A \kappa^\mathrm{c} / k^\mathrm{s}$. 
This result predicts that the local u-shaped bolalipid fraction $u_\mathrm{f}$ increases linearly with the squared mean curvature $H^2$, consistent with previous observations \cite{amaral2023}.
The prediction can be confirmed by correlating the u-shaped bolalipids fraction $u_\mathrm{f}$ with the reduced squared mean curvature $H^2 \cdot \rcross^2$, using the corresponding toroidal aspect ratio $\rho$ (Fig.~\ref{fig:Figure4}C).
We conclude that membrane shape, and in particular the squared mean curvature, both locally and globally enforces the enrichment of u-shaped bolalipids.

We next examine our simulations of archaeal mixture membranes, in which u-shaped bolalipids are rare and bilayer lipids confer membranes flexible. 
We ask whether bilayer lipids show local enrichment along the membrane surface, analogous to the curvature-driven enrichment of u-shaped bolalipids.
To address this, we focus on mixture membranes with bilayer lipid fractions $\fbi = 0.7-0.9$ and relatively stiff bolalipids ($\km=\unit[2]{}\kBT$), where the toroidal shape persists throughout the simulation trajectories (Fig.~\ref{fig:Figure3}A-C).
The stability of the toroidal geometry is further confirmed by analyzing the ring radius $\Rring$, the cross-section radius $\rcross$, the toroidal aspect ratio $\rho$ and the hole radius $\Rring - \rcross$ for $\fbi=0.7$ (Fig.~S4), $\fbi=0.8$ (Fig.~S5) and $\fbi=0.9$ (Fig.~S6), which vary only weakly over time.
We conclude that the toroidal shape is stable, allowing us to quantify the local fraction of the bilayer lipids along the torus surface.

Following the same protocol used for u-shaped bolalipids, we determine the local bilayer fraction $\fbi$, averaged over $N_\mathrm{seeds}=4$ independent trajectories, along the membrane surface.
In Fig.~\ref{fig:Figure4}D, the local bilayer fraction for $\fbi=0.7$ is plotted as a function of the toroidal angle $\vartheta$ (black line) with standard error (gray shaded area).
We fit the scaled squared mean curvature $\tilde{H^2}$ (red line) and scaled Gaussian curvature $\tilde{K}$ (purple line) of the Clifford torus to the data.
Fig.~\ref{fig:Figure4}D shows that the bilayer fraction $\fbi$ is not uniform along the torus surface but instead correlates with the squared mean curvature.
This trend is consistent for Clifford tori with higher bilayer fractions, $\fbi=0.8$ and $\fbi=0.9$ (Fig.~S7).
From the fits, we extract values of the toroidal aspect ratio $\rho$ and compare them to the geometrical analysis of the torus shape in Figs.~S4-6. 
The small deviations likely arise from shape variations and irregularities of the membrane tori (Table~\ref{table:rho}).
Overall, the agreement confirms that the membrane geometry governs the local distribution of both bilayer and bolalipids in mixture membranes.

Based on Eq.~(\ref{eq:uf_with_H2}),
we also expect that the local bilayer fraction $\fbi$ increases linearly with the squared mean curvature $H^2$ in mixture membranes.
This expectation is confirmed when plotting the bilayer fraction $\fbi$ as a function of the reduced squared mean curvature $H^2 \cdot \rcross^2$, using the corresponding toroidal aspect ratio $\rho$ for $\fbi=0.7$ (Fig.~\ref{fig:Figure4}E and Fig.~S8A).
Similar linear correlations are observed for Clifford tori with higher bilayer fractions, $\fbi=0.8$ (Fig.~S8B) and $\fbi=0.9$ (Fig.~S8C).

To quantify the effect of the overall bilayer lipid fraction on local lipid organization, we define the bilayer heterogeneity as $(\fbi^\mathrm{max}-\fbi^\mathrm{min})/\fbi^\mathrm{max}$, where $\fbi^\mathrm{max}$ and $\fbi^\mathrm{min}$ are the maximum and minimum local bilayer fractions along the toroidal angle $\vartheta$.
Fig.~\ref{fig:Figure4}F shows that bilayer heterogeneity decreases with increasing global bilayer fraction $\fbi$, reflecting that the relative enrichment of bilayer lipids along high-curvature regions becomes less pronounced as more bilayer lipids are available.

\section{Discussion}
In this work, we investigated the interplay between lipid geometry, membrane curvature, shape stability and lipid organization in archaeal membranes.   
To probe curvature effects on membranes, a variety of approaches have previously been employed in simulations and experiments, including scaffolding particles \cite{konig2023}, adhering cargo beads \cite{cooke2006,amaral2023} and pulling membrane tethers \cite{sorre2009,baoukina2018}.
Here, we assembled membranes composed of bilayer and bolalipids into the shape of a Clifford toroidal vesicle, which allowed us to systematically disentangle the effects of mean and Gaussian curvature within a single membrane geometry.
Our coarse-grained molecular dynamics simulations access regimes of large curvature, as the simulated toroidal vesicles have diameters of about $\unit[70]{nm}$.
This complements earlier experimental studies of toroidal vesicles formed from polymerized membranes at the micrometer scale \cite{mutz1991}.
Moreover, the Clifford torus represents a metastable geometry of the membrane shape energy and can undergo shape transitions.
Therefore, our approach allowed us to investigate how membrane structure and composition jointly control membrane shape stability and lipid organization. 

For pure bolalipid membranes, we found that the bolalipid rigidity $\km$ governs membrane shape stability.
For membranes at vanishing bolalipid rigidity ($\km=0$), containing approximately 50\% u-shaped bolalipids and exhibiting a bending rigidity of $\kappa\approx\unit[8]{}\kBT$ \cite{amaral2023}, the toroidal vesicle remained stable.
As the bolalipid rigidity $\km$ increased, most bolalipids adopted a straight conformation, the membrane approached a monolayer geometry and the bending rigidity $\kappa$ increased.
In this regime, the toroidal vesicle either transformed into a spherical vesicle or ruptured and ultimately became flat.
For mixture membranes composed of bilayer lipids and rigid bolalipids, we found that the bilayer fraction controls membrane shape and stability.
At small bilayer fractions, corresponding to a large membrane rigidity, the toroidal vesicles were unstable and transitioned into spherical vesicles. 
In contrast, larger bilayer fractions and thus smaller bending rigidities, stabilized the toroidal vesicle, preventing further transitions.

A comparison between pure bolalipid membranes and mixed membranes revealed a qualitative difference in the torus-to-sphere transition pathway. 
In mixed membranes at low bilayer fraction $f_\mathrm{bi}=0.1$, the transition was accompanied by transient pores, consistent with the large bending rigidity  of $\kappa\approx\unit[50]{}\kBT$.
By contrast, in pure bolalipid membranes at a moderate bolalipid rigidity $\km=\unit[1]{}\kBT$, the membrane deformed smoothly without pore formation, reflecting a lower rigidity of $\kappa\approx\unit[30]{}\kBT$ \cite{amaral2023}.

While curvature-induced lipid sorting has been extensively investigated for bilayer membranes in both experiments and simulations \cite{cooke2006,tian2009,kamal2009,sorre2009,beltran-heredia2019,konig2023,pohnl2023}, it has not been previously studied for bolalipid membranes or bolalipid-bilayer mixtures.
To elucidate how membrane curvature affects lipid organization in these systems, we analyzed the spatial distribution of lipids in toroidal vesicles formed from archaeal membranes. 
We found that both u-shaped bolalipids and bilayer lipids preferentially accumulate in regions of high squared mean curvature, independent of Gaussian curvature.
This curvature-driven sorting emerges from a balance between minimizing the membrane bending energy and maximizing the mixing entropy of the lipids.
Despite being modest in magnitude ($\sim 1.5\%$ for u-shaped bolalipids, $\sim 5\%$ for bilayer lipids), the enrichment identifies curvature-composition coupling as a physical signature of archaeal membrane remodeling.

Our work suggests an interesting experimental setup.
Since rigid bolalipids are depleted from regions of high membrane curvature, whereas u-shaped bolalipids preferentially accumulate in such regions, both species could serve as intrinsic membrane curvature sensors.  
By fluorescently labeling bolalipids, local membrane curvature could be inferred from spatial variations in fluorescence intensity, enabling direct visualization and quantitative analysis of curvature distributions.

Our results also hint at an important distinction between bolalipid and bilayer membranes. 
While bolalipids must undergo conformational change to transition from the straight to the u-shaped conformation in order to accommodate regions of high membrane curvatures, bilayer lipids can simplify diffuse within the membrane plane to such regions.
In our model, the flip-flop rates of bilayer lipids and thus also the rates of conformational change of bolalipids are likely higher than in biological membranes \cite{cooke2005,cooke2006}.
As a consequence, our simulations reach steady state membrane configurations  relatively rapidly through a combination of diffusion and lipid conformational changes,  leading to similar membrane remodeling dynamics in membranes composed of u-shaped bolalipids and bilayer lipids.
In biological membranes, however, the relevant timescales differ substantially. 
While proteins that promote lipid flip-flop events, such as flippases and scramblases have been identified also in archaea \cite{makarova2015,verchere2017}, it remains an open question whether analogous proteins exist that catalyze conformational transitions of bolalipids from straight to u-shaped states \cite{bhattacharya2024}. 
This disparity implies that, although bolalipid membranes may be as flexible as bilayer membranes, the dynamics of shape transformations in bolalipid membranes could be much slower than in bilayer membranes in the absence of proteins that facilitate the transition from straight to u-shaped conformations.

Our study provides insight into fundamental biological processes in the archaeal branch of the tree of life, such as cell division and vesicle formation, which all induce dramatic curvature changes. 
Moreover, our work reveals curvature-composition coupling as a hallmark of archaeal membrane remodeling.

\section{Data availability}
The simulation input files are freely available at https://zenodo.org/records/18772087.

\section{Supplementary material}
Details on the computational model and the simulation analysis, including Figs. S1-S8 and Videos S1 and S2, are available in the supplementary material. 

\section{Acknowledgments}
FF acknowledges financial support by the NOMIS foundation.
MA and A\v{S} acknowledge funding by the Volkswagen Foundation Grant Az 96727.
A\v{S} acknowledges funding by ERC Starting Grant “NEPA” 802960 and Vallee scholarship.

\bibliographystyle{apsrev4-2}
\bibliography{Bibliography}

\newgeometry{onecolumn}
\onecolumngrid

\begin{center}
\textbf{\large Supplementary material\\
Cracking donuts and sorting lipids: geometry controls archaeal membrane stability and lipid organization} \\
Felix Frey, Miguel Amaral, Anđela Šarić
\end{center}

\setcounter{equation}{0}
\setcounter{figure}{0}
\setcounter{table}{0}
\setcounter{page}{1}
\setcounter{section}{0}
\makeatletter
\renewcommand{\thefigure}{S\arabic{figure}}
\renewcommand{\theequation}{S\arabic{equation}}

\section{Computational model}
\subsection{Coarse-grained particle based model of bilayer and bolalipids}
To model bilayer and bolalipids we follow the model of Cooke and Deserno for bilayer lipids \cite{cooke2005} and our extension to bolalipids \cite{amaral2023}. 
In the model, a bilayer lipid consists of one head bead and two tail beads (Fig.~1B).
A bolalipid is represented by joining two bilayer lipids along their tails, so it consists of two head beads that are connected by four tail beads (Fig.~1B).
As a result, bilayer and bolalipids contain identical head and tail groups.
Adjacent beads of bilayer and bolalipids are connected by finite extensible nonlinear elastic (FENE) bonds.
Including a repulsive Lennard-Jones potential that enforces volume exclusion, we obtain the following potential:
\begin{align}
U_\mathrm{bond}=
\begin{cases}
-\frac{1}{2} K R_0^2 \ln \left[1- \left( \frac{r}{R_0}\right)^2 \right]
+\epsilon \left[ 
\left(
\frac{\rmm}{r}
\right)^{12}
-2
\left(
\frac{\rmm}{r}
\right)^{6}
+1\right]
  \, , & 0\le  r < r_\mathrm{c}, \\
-\frac{1}{2} K R_0^2 \ln \left[1- \left( \frac{r}{R_0}\right)^2 \right]
\, , & r_\mathrm{c} \le r \le R_0 .
\end{cases}
\label{eq:FENE}
\end{align}

The first term of the potential is attractive, while the second term is repulsive.
In the first term, describing the bond, 
$K=\unit[30]{}\kBT \sigma^{-2}$ is the energy density of the bond and $R_0=\unit[1.5]{\sigma}$ defines the maximum extent of the bond.
In the second term, describing volume exclusion,
$\epsilon=\unit[1]{}\kBT$ defines the energy scale of the repulsive interaction with the potential minimum at $\rmm=\unit[2^{\frac{1}{6}}]{\sigma}\approx \unit[1.122]{\sigma}$ and the cut-off at the minimum of the potential $r_\mathrm{c}=\rmm$.
We note that in the simulations our time, length and energy units are $t_0$, $\sigma$ and $\kBT$ with the Boltzmann constant $k_\mathrm{B}=1$. As a result, the unit of mass is given by $1m=1\kBT t_0^{2} \sigma^{-2}$.

Between the first, second and third bead of a bilayer lipid we define the angle $\alpha$ for which we introduce a harmonic angle potential of the following form
\begin{equation}
U_\mathrm{angle}=K_\alpha \left(\alpha-\alpha_0 \right)^2 \, ,
\label{eq:angular_potential}
\end{equation}
with $K_\alpha=k_0=\unit[5]{}\kBT$ and $\alpha_0=\pi$.
For a bolalipid, we introduce identical harmonic angle potentials between the first three beads and the last three beads with $K_\alpha=k_0$.
In addition, we introduce two angle potentials between the second, third and fourth bead and between the third, fourth and fifth bead for bolalipids with $K_\alpha=\km$, varying between $\unit[0]{}\kBT$ and $\unit[5]{}\kBT$.

Head beads of different lipids interact with each other and with non-bonded tail beads solely through volume exclusion. 
The repulsive potential is given by

\begin{align}
U_\mathrm{head\,bead}=
\begin{cases}
 \epsilon \left[ 
\left(
\frac{\rmm}{r}
\right)^{12}
-2
\left(
\frac{\rmm}{r}
\right)^{6}
+1\right]
\, , & 0\le  r < r_\mathrm{c}, \\
0
\, , & r_\mathrm{c} \le r .
\end{cases}
\label{eq:volume}
\end{align}

In Eq.~(\ref{eq:volume}), both the minimum of the potential and the cut-off are set to $\rmm=r_\mathrm{c}=\unit[0.95\cdot 2^{\frac{1}{6}}]{\sigma}$, similar to \cite{cooke2005}.
For bilayer lipids, the energy scale is given by $\epsilon=\unit[1/1.3]{}\kBT$.
In contrast, to represent bolalipids at different molecular rigidities $\km$ and to keep the assembled membranes in a fluid state, we follow our previous approach and vary the energy scale in Eq.~(\ref{eq:volume}) given by $\epsilon=\epsilon_\mathrm{p}=1/(1.3+\km/(10\,\kBT))\kBT$ \cite{amaral2023}.

Non-bonded tail beads interact with each other according to the following potential: 

\begin{align}
U_\mathrm{tail\,bead}=
\begin{cases}
-\epsilon+\epsilon \left[ 
\left(
\frac{\rmm}{r}
\right)^{12}
-2
\left(
\frac{\rmm}{r}
\right)^{6}
+1\right]
\, , &  0\le  r < \rmm,\\
- \epsilon \cos^2 \left( \frac{\pi (r-\rmm)}{2\omega}\right)
\, , & \rmm \le r < r_\mathrm{c}, \\
0
\, , & r_\mathrm{c}\le r .
\end{cases}
\label{eq:attraction}
\end{align}

In Eq.~(\ref{eq:attraction}), the bilayer tail-tail interaction energy is given by $\epsilon=\unit[1/1.3]{}\kBT$, the potential minimum is at $\rmm=\unit[2^{\frac{1}{6}}]{\sigma}$ and the cut-off at $r_\mathrm{c}=\rmm+\omega$, where $\omega=\unit[1.5]{\sigma}$ is the range of the interaction potential.
For bolalipid tail bead interactions, we use $\epsilon_\mathrm{p}=1/(1.3+\km/(10\,\kBT))\kBT$.
When we simulate mixture membranes of bilayer lipids and bolalipids we use $\km=\unit[2]{}\kBT$ and $\epsilon=\unit[1/1.3]{}\kBT$ so that bilayer lipids and bolalipids interact in the same way. 

\subsection{Initial distribution of lipids on the surface of a torus}
In our simulations, we initially distribute lipids along a Clifford torus. 
We therefore randomly sample lipid positions from a torus geometry with ring radius $\Rring=\unit[\sqrt{2}\cdot 15]{\sigma}$, cross section radius $\rcross=\unit[15]{\sigma}$ and toroidal aspect ratio $\rho=\Rring/\rcross=\sqrt{2}$.
Depending on whether we simulate a bolalipid or mixture membrane, we either place one bolalipid or two bilayer lipids lined up in radial direction at each sampled position. 
In Table~\ref{table:lipids} we summarize the number of bolalipids ($\#$ bolalipids), bilayer lipids ($\#$ bilayer lipids), the total number of lipids ($\#$ lipids) and the lipid bead number ($\#$ beads) used for different values of the bilayer fraction $\fbi$. 

\begin{table}[t]
\caption{The used number of bolalipids ($\#$ bolalipids), bilayer lipids ($\#$ bilayer lipids), the total number of lipids ($\#$ lipids) and the lipid bead number ($\#$ beads) for different values of the bilayer fraction $\fbi$.}
\label{table:lipids}
\begin{tabular}{ccccccccccc}
\hline
$\fbi$ & 0.0 (pure bolalipid membrane) &  0.1 &  0.2 &  0.3 &  0.4 &  0.5 &  0.6 &  0.7 &  0.8 &  0.9  \\
\hline
$\#$ bolalipids & 9000 &  8526 &  8000 &  7411 &  6750 &  6000 &  5142 &  4153 &  3000 &  1638  \\
$\#$ bilayer lipids & 0 &  946 &  2000 &  3176 &  4500 &  6000 &  7714 &  9692 &  12000 &  14726  \\
$\#$ lipids & 9000 &  9472 &  10000 &  10589 &  11250 &  12000 &  12856 &  13845 &  15000 & 16364   \\
$\#$ beads & 54000 &  53994 &  54000 &  53994 &  54000 &  54000 &  53994 &  53994 &  54000 & 53994   \\
\hline
\end{tabular}
\end{table} 

\subsection{Relaxation phase of toroidal membrane vesicles and simulation geometry}

After the lipids have been distributed along the torus geometry, the lipid positions are relaxed to avoid overlapping lipids. 
To accommodate lipid position relaxation, we implement a soft potential that acts between all pairs of beads of different lipids and is parametrized as 
\begin{equation}
U_\mathrm{soft}=A \left [1+\cos \left ( \frac{\pi r}{\rc}\right ) \right] \, ,
\end{equation}
where $A$ is the energy scale of the potential and $\rc$ is the cut-off.
In the simulations, we initially relax the lipid bead positions for $3\cdot 10^3 \ts$ timesteps with a timestep of $t_\mathrm{s}=\unit[0.1]{t_0}$.
We increase $A$ from $\unit[3]{}\kBT$ to $\unit[7]{}\kBT$ in steps of $\unit[2]{}\kBT$ and $\rc$ from $\unit[0.7]{\sigma}$ to $\unit[1]{\sigma}$  in steps of $\unit[0.2]{\sigma}$ and $\unit[0.1]{\sigma}$ every $10^3$ timesteps.

We use a simulation box with periodic boundary conditions of dimensions $L_\mathrm{x}=\unit[100]{\sigma}$, $L_\mathrm{y}=\unit[100]{\sigma}$ and $L_\mathrm{z}=\unit[100]{\sigma}$.
To simulate the system, we use the LAMMPS MD simulation package \cite{thompson2022}.
During the relaxation phase, we time integrate the positions and velocities of all lipid beads, carrying a mass of $1m=1\kBT \, t_0^2 / \sigma^2$ each.
We limit the maximum distance a lipid bead can move in one timestep to $\unit[0.1]{\sigma}$ between the first $2\cdot10^3$ timesteps and then to $\unit[0.01]{\sigma}$ during the following $10^3$ timesteps, implemented in LAMMPS via fix nve/limit.
To visualize the system we use OVITO \cite{stukowski2009}.

\subsection{Integrating equations of motion}
After the lipid positions have been sufficiently relaxed we start the simulation with the model as detailed in Eqs.~(\ref{eq:FENE}-\ref{eq:attraction}).
We simulate the system using overdamped Langevin dynamics in a fixed volume, implemented in LAMMPS via fix nve + fix langevin.
We set the damping coefficient of the thermostat $\gamma=1 \, t_0$ and use a timestep of $\ts = 0.01 t_0$.
We run the simulations for $4\cdot 10^6$ timesteps and save the bead positions every $2\cdot 10^3$ timesteps.
For the different membranes, we typically use $N_\mathrm{seeds}=4$ random initial seeds. For pure bolalipid membranes at $\km=\unit[0]{}\kBT$, we use $N_\mathrm{seeds}=16$ random initial seeds to increase the statistics.

\section{Simulation analysis}
\subsection{Measurement of the reduced volume}
\textbf{Pure bolalipid membranes}
In order to measure the reduced volume of pure bolalipid membranes we first build a surface mesh with Ovito from the set of lipid beads using the alpha-shape method with radius $\unit[2]{\sigma}$ and smoothing level 10.
From the surface mesh we compute the surface area $A$ and the enclosed volume $V$, which allows us, using Eq.~(5), to calculate the reduced volume $\nu$.
We compute $\nu$ along the trajectory every $2\cdot10^4$ simulation steps. 
For each molecular rigidity $\km$, we average $\nu$ for $N_\mathrm{seeds}=4$ seeds and calculate the mean and standard error of the mean. 
We note that for some time points along the trajectory the construction of the surface mesh fails ($\nu=0$), which we ignore when calculating the average of $\nu$. 
Also when transient membrane pores start to open ($\km\le\unit[1]{}\kBT$), $\nu$ cannot be computed.
For $\km>\unit[1]{}\kBT$ the toroidal vesicles break within $2\cdot 10^5$ timesteps, which is why we do not compute $\nu$ in this case.
The results of the analysis are shown in Fig.~2B.

\textbf{Membrane mixtures of bilayer and bolalipids} We repeat the procedure as outlined for pure bolalipid membranes.
The results of the analysis are shown in Fig.~3B.
For $\fbi=0.1$, the mean of $\nu$ cannot be computed along the full trajectory because the membrane breaks and forms transient pores.

\subsection{Theoretical value of the reduced volume for the Clifford torus}
In this section, we calculate the theoretical value of the reduced volume $\nu_\mathrm{Clifford}$ for the Clifford torus.
The volume and surface area of a torus are given by 
\begin{align}
V_\mathrm{torus}&=2 \pi^2 \rcross^2 \Rring \, ,
\\
A_\mathrm{torus}&=4 \pi^2 \Rring \rcross \, .
\end{align}
Thus, for the Clifford torus with $\Rring/\rcross=\sqrt{2}$,
the reduced volume according to Eq.~(5) is given by 
\begin{align}
\nu_\mathrm{Clifford}=\frac{3 V_\mathrm{Clifford}}{4\pi  \left(\frac{A_\mathrm{Clifford}}{4\pi}\right)^{3/2}}
= \frac{3}{2} \sqrt{\frac{\rcross}{\pi \Rring}}= \frac{3}{2} \sqrt{\frac{1}{\pi \sqrt{2}}} 
\approx 0.712 \, .
\end{align}
The result is shown in Figs.~2B and 3B.
The observed small deviation of the reduced volume $\nu$ for the simulated torus from the theoretical value $\nu_\mathrm{Clifford}$ is expected due to finite thickness of the membrane. 
While in the calculation the Clifford torus is parametrized for the mid-plane of the membrane, in the simulation the enclosed volume is calculated from the outer surface mesh, effectively increasing the reduced volume.

\subsection{Quantifying the number of membrane handles or pores}
To measure the number of membrane handles or pores, we use the surface mesh constructed in Ovito as described above.
Because of the finite thickness of the simulated membranes, we reconstruct two surface meshes for a closed toroidal vesicle that represent the outer and inner surface layer of the membrane.
As soon as the toroidal vesicle breaks, we only reconstruct a single surface mesh.
The surface mesh is given by the number of vertices $V$, edges $E$ and faces $F$.
Following Euler's polyhedron formula, we can calculate the genus $g$ of the mesh by $g=N-(V+F-E)/2$, where $N$ is the number of surfaces.

Since we can have either two surface meshes (as long as the membrane has not ruptured) or one in our simulations, the interpretation of $g$ depends on $\nu$.
For $\nu>0$, the genus $g$ represents the number of handles of the membrane. 
In this case, we correct for the fact that the calculation is performed on two surface meshes and divide the obtained $g$ by 2.
Thus, we expect $g=1$ for a torus and $g=0$ for a sphere. 
If $\nu$ can no longer be computed because the membrane is no longer closed, 
$g$ denotes the number of pores perforating the membrane.
Thus, we expect $g=n$ for an open surface containing $n$ pores.

We compute $g$ throughout the simulation trajectories every $2\cdot10^4$ simulation steps. 
For each molecular rigidity $\km$ or bilayer fraction $\fbi$, we average $g$ over $N_\mathrm{seeds}=4$ independent seeds and calculate the mean and standard error of the mean. 
The results of the analysis are shown in Fig.~2C and Fig.~3C.

\begin{figure}[t]
\centering
\includegraphics[width=.95\textwidth]{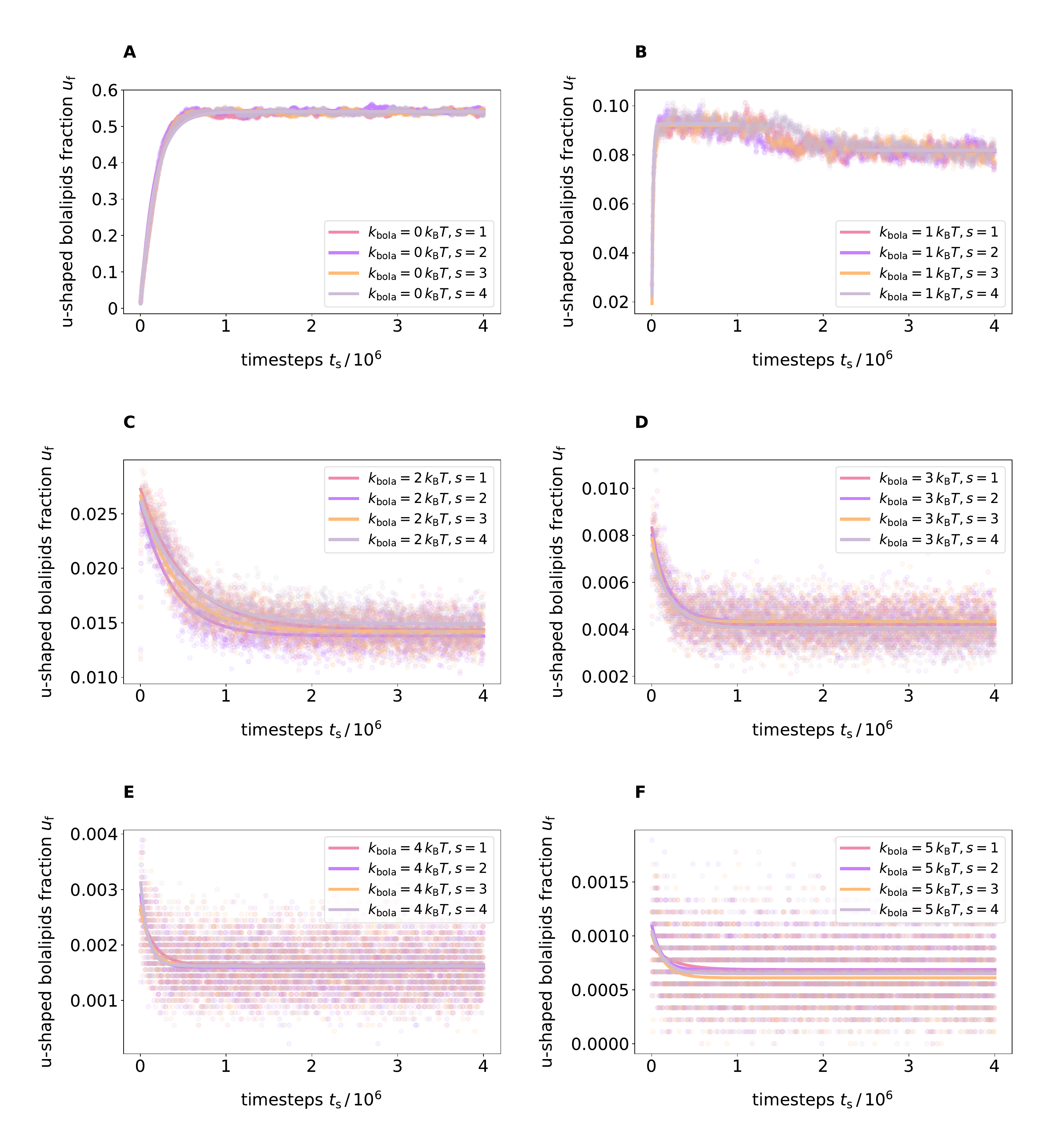}
\caption{Fraction of u-shaped bolalipids as a function of time for increasing bolalipid rigidity $\km$ (A-F). 
From fits to the data (see text), we determine the steady state value of the fraction of u-shaped bolalipids $\uf^0$. 
We average the values of $\uf^0$ for the different seeds and show the results in Fig.~2D.
Data shown is based on $N_\mathrm{seeds}=4$ random initial seeds.
\label{fig:FigureS1}}
\end{figure}

\subsection{Global fraction of u-shaped bolalipids}
\textbf{Pure bolalipid membranes} 
To measure the confirmation of bolalipids, we evaluate the angle $\theta$ between vectors that connect the central beads and the head beads for every simulation timestep.
Based on $\theta$, we classify bolalipids as straight ($\theta<\pi/2$) or u-shaped ($\theta\ge\pi/2$).
By averaging over all lipids, we calculate the global fraction of u-shaped bolalipids $\uf$ for one timestep.
Fig.~\ref{fig:FigureS1} shows $\uf$ as a function of time for different values of $\km$.
In all cases, we find similar results for independent simulation seeds.

For $\km=\unit[0]{}\kBT$, the u-shaped bolalipid fraction $\uf$ rapidly increases and then saturates (Fig.~S1A).
To determine the saturation level for each of the seeds, we fit an exponential function
$\uf^1(t)=\uf^0 (1-\exp(-t/\tau))$, with the saturation level $\uf^0$ and the saturation time scale $\tau$.
We average the different values for $\uf^0$ and plot the result in Fig.~2D.
For $\km=\unit[1]{}\kBT$ (Fig.~\ref{fig:FigureS1}B), $\uf$ exhibits an initial increase followed by saturation, after which it decreases before reaching its final saturation state.
The decrease is associated with the transition from torus shape to sphere (Fig.~2A).
This observation suggests that membrane shape and specifically membrane curvature globally enforce u-shaped bolalipids.
We fit a constant $\uf^2(t)=\uf^0$ to the second saturation phase (cf.~Fig.~\ref{fig:FigureS1}B), average $\uf^0$ over the different seeds and plot the result in Fig.~2D.
For $\km\ge\unit[2]{}\kBT$ (Fig.~\ref{fig:FigureS1}C-F), we observe that $\uf$ slightly increases in the initial toroidal state and subsequently decreases as the membranes breaks and becomes flat.
We fit a relaxation function $\uf^3(t)=\uf^i \exp(-t/\tau)+\uf^0$, where $\uf^i$ is the initial fraction of u-shaped bolalipids, average $\uf^0$ over  different seeds and plot the result in Fig.~2D.

We compare our results with previous measurements for the u-shaped bolalipid fraction $\uf$ for flat membranes \cite{amaral2023}. 
It was found that $\uf(\km)=1/(1+\exp (\beta(-0.16\kBT+3\km)))$, where $\beta=1/(\kBT)$, describes the data.
We note that in this case, a value of $\epsilon_\mathrm{p}=1/1.2\kBT$ was used for $\km=0$.
The result is shown in Fig.~2D as a gray line.

\begin{figure}[t]
\centering
\includegraphics[width=.95\textwidth]{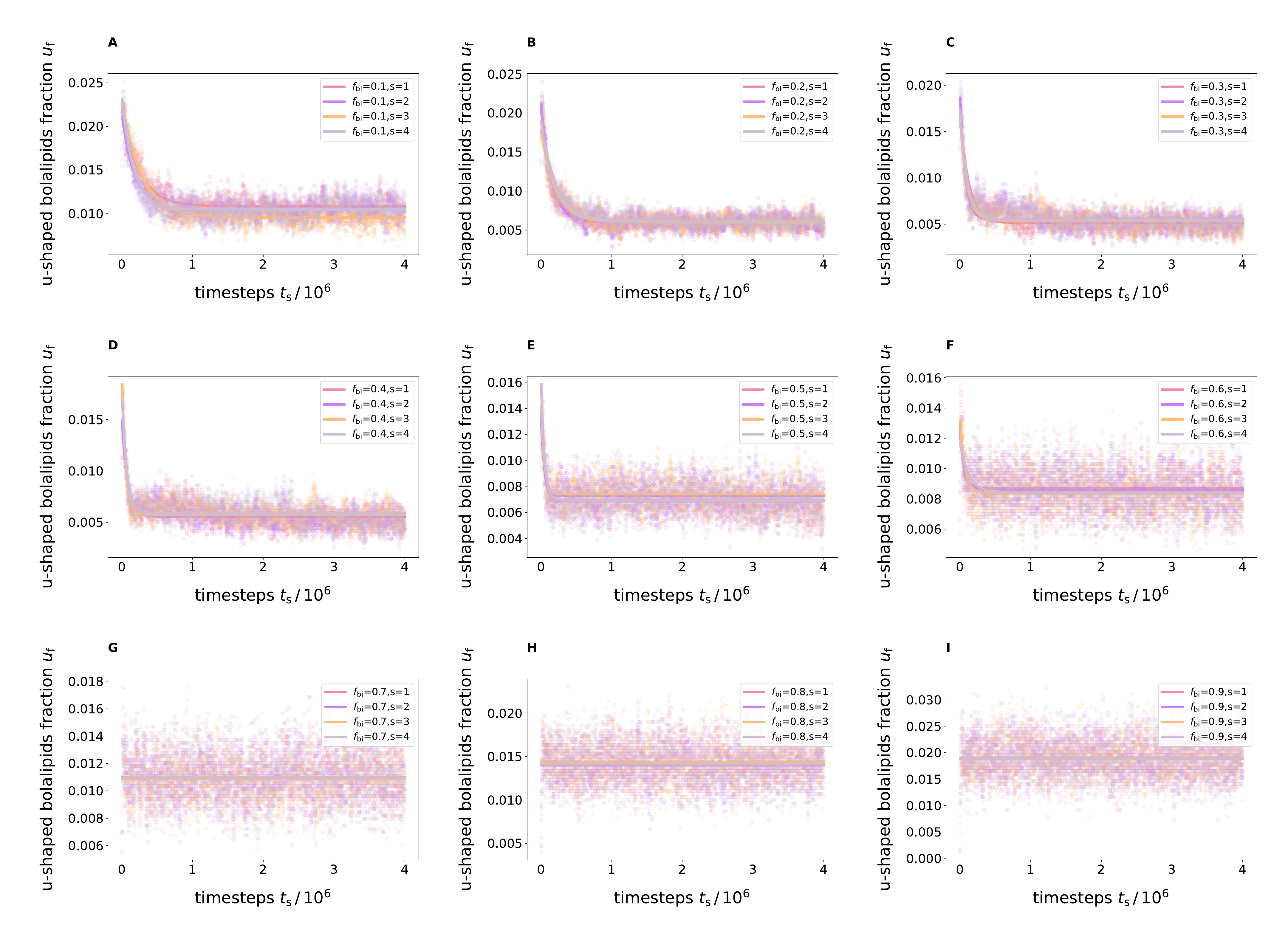}
\caption{
Fraction of u-shaped bolalipids as a function of time for increasing bilayer fraction $\fbi$ (A-I).
From fits to the data (see text), we determine the steady state value of the fraction of u-shaped bolalipids $\uf^0$. 
We average the values of $\uf^0$ for the different seeds and show the results in Fig.~3D.
Data shown is based on $N_\mathrm{seeds}=4$ random initial seeds.
\label{fig:FigureS2}}
\end{figure}

\textbf{Membrane mixtures of bilayer and bolalipids}
To measure the global fraction of u-shaped bolalipids for mixture membranes, we plot $\uf$ as a function of time in Fig.~\ref{fig:FigureS2}.
For $\fbi=0.1-0.6$ (Fig.~\ref{fig:FigureS2}A-F) we use $\uf^3(t)$ and for $\fbi=0.7-0.9$ (Fig.~\ref{fig:FigureS2}G-I) we use $\uf^2(t)$ to fit the data.
We average $\uf^0$ over the different seeds and plot the result in Fig.~3D.

\begin{figure}[t]
\centering
\includegraphics[width=.95\textwidth]{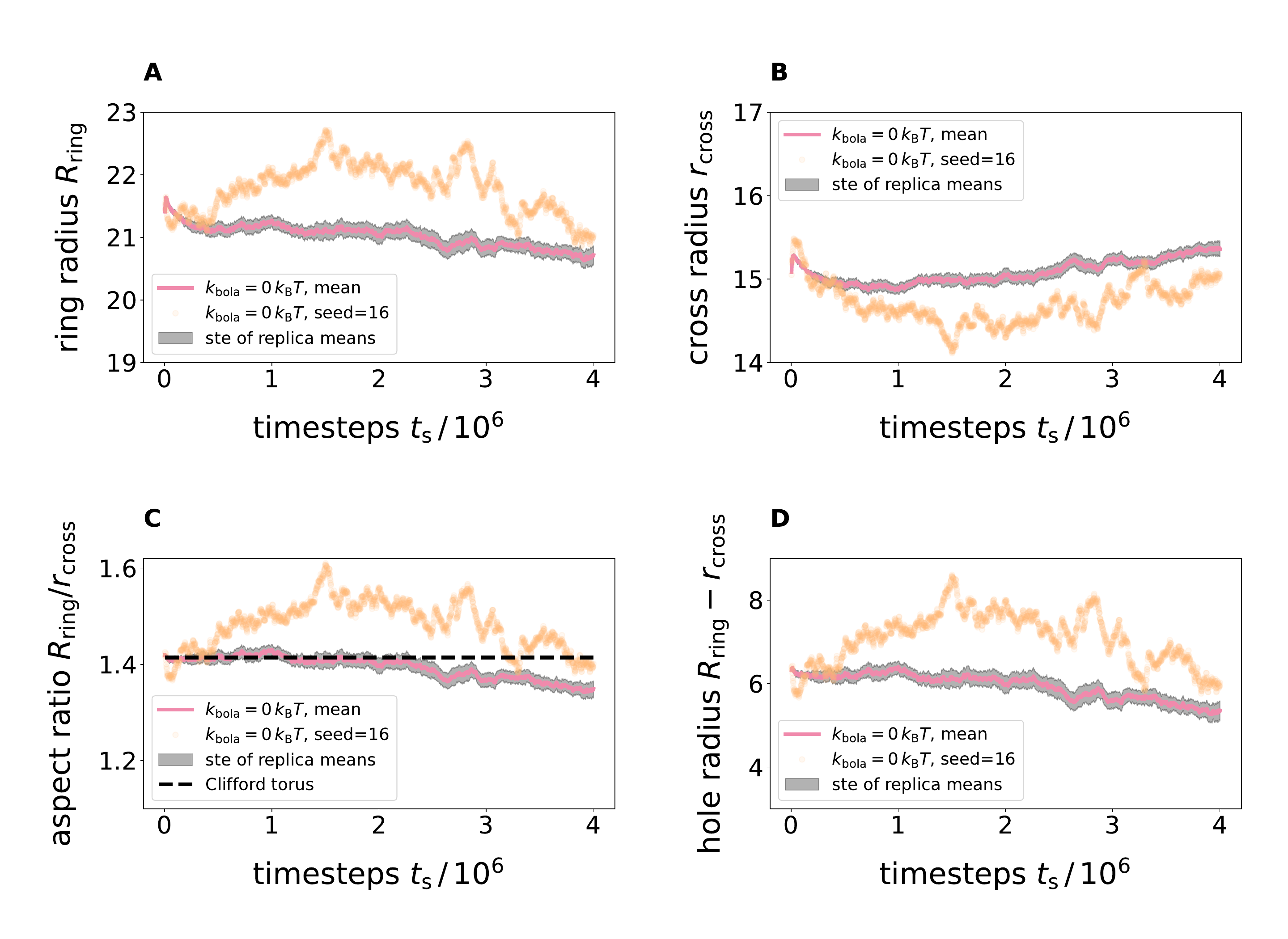}
\caption{
Geometrical analysis of toroidal vesicles of pure bolalipid membranes at $\km=\unit[0]{}\kBT$.
(A) Ring radius as a function of time. 
(B) Cross section radius as a function of time.
(C) Toroidal aspect ratio as a function of time.
(D) Hole radius as a function of time.
Data shown is based on $N_\mathrm{seeds}=16$ random initial seeds.
\label{fig:FigureS3}}
\end{figure}

\subsection{Analysis of the geometry of toroidal vesicles}

In this section, we explain how we analyze the geometry of toroidal vesicles.
First, we extract the positions of all membrane beads for each timestep of the simulation, which we center around the center of mass.
We then map the Cartesian coordinates $(x,y,z)$ of the membrane beads to spherical coordinates with radius $R=\sqrt{x^2+y^2+z^2}$ and angle $\theta=\arcsin(z/R)$.
Assuming rotational symmetry around the z-axis, we neglect to map the polar angle $\phi$. 
Then, all membrane bead coordinates are given in the $\chi z-$plane with $\chi =R \cos \theta$ and $z$.
We minimize the function $f(R, \theta;\Rring, \rcross)=\rcross^2-(\Rring-\chi)^2-z$ to determine the ring radius $\Rring$ and the cross section radius $\rcross$.
We compute $\Rring$ and $\rcross$ along the trajectories every $2\cdot10^3$ simulation steps. 

\begin{figure}[t]
\centering
\includegraphics[width=.95\textwidth]{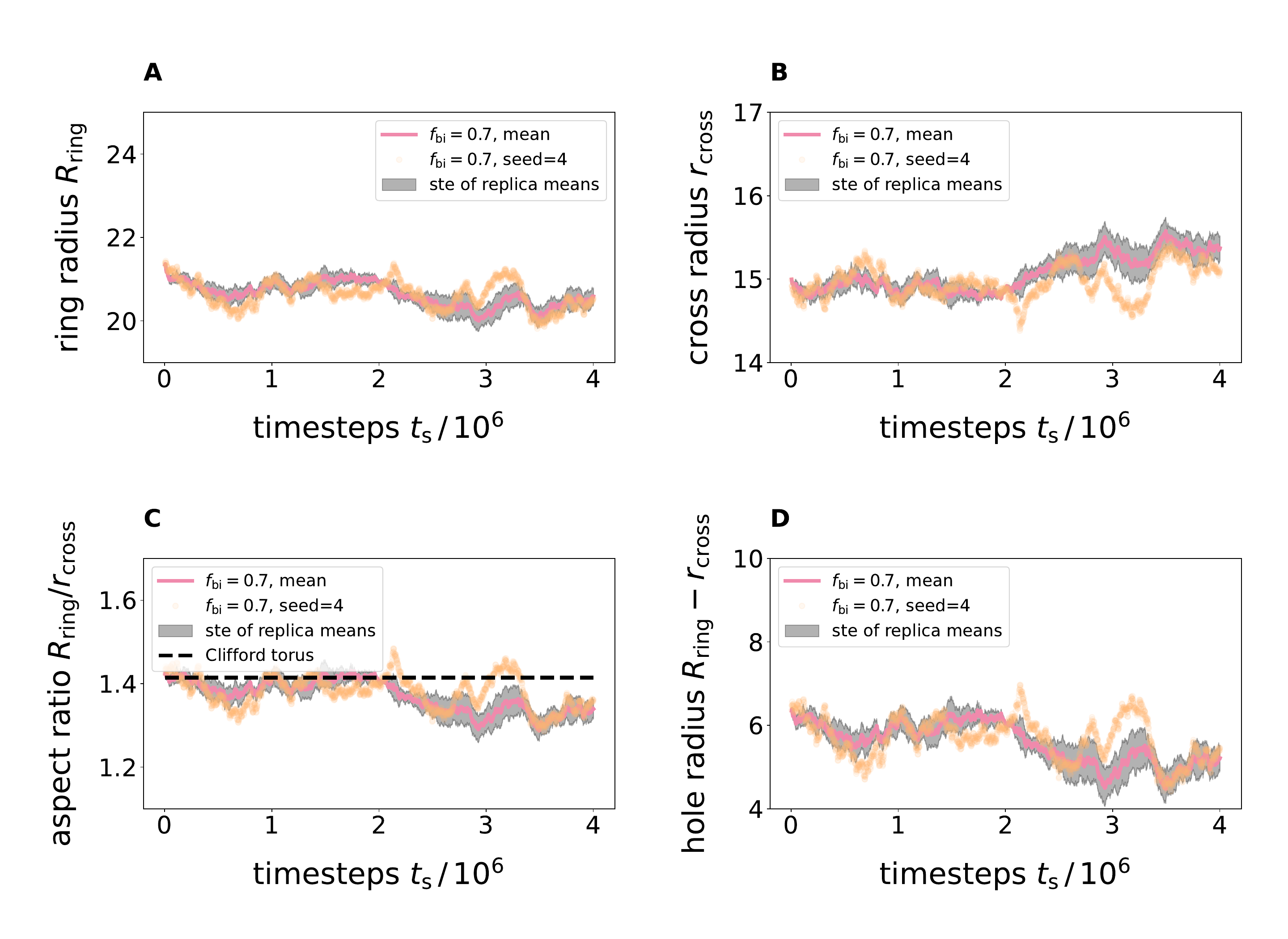}
\caption{Geometrical analysis of toroidal vesicles of mixed membranes at $\fbi=0.7$.
(A) Ring radius as a function of time. 
(B) Cross section radius as a function of time.
(C) Toroidal aspect ratio as a function of time.
(D) Hole radius as a function of time.
Data shown is based on $N_\mathrm{seeds}=4$ random initial seeds.
\label{fig:FigureS4}}
\end{figure}

\begin{figure}[t]
\centering
\includegraphics[width=.95\textwidth]{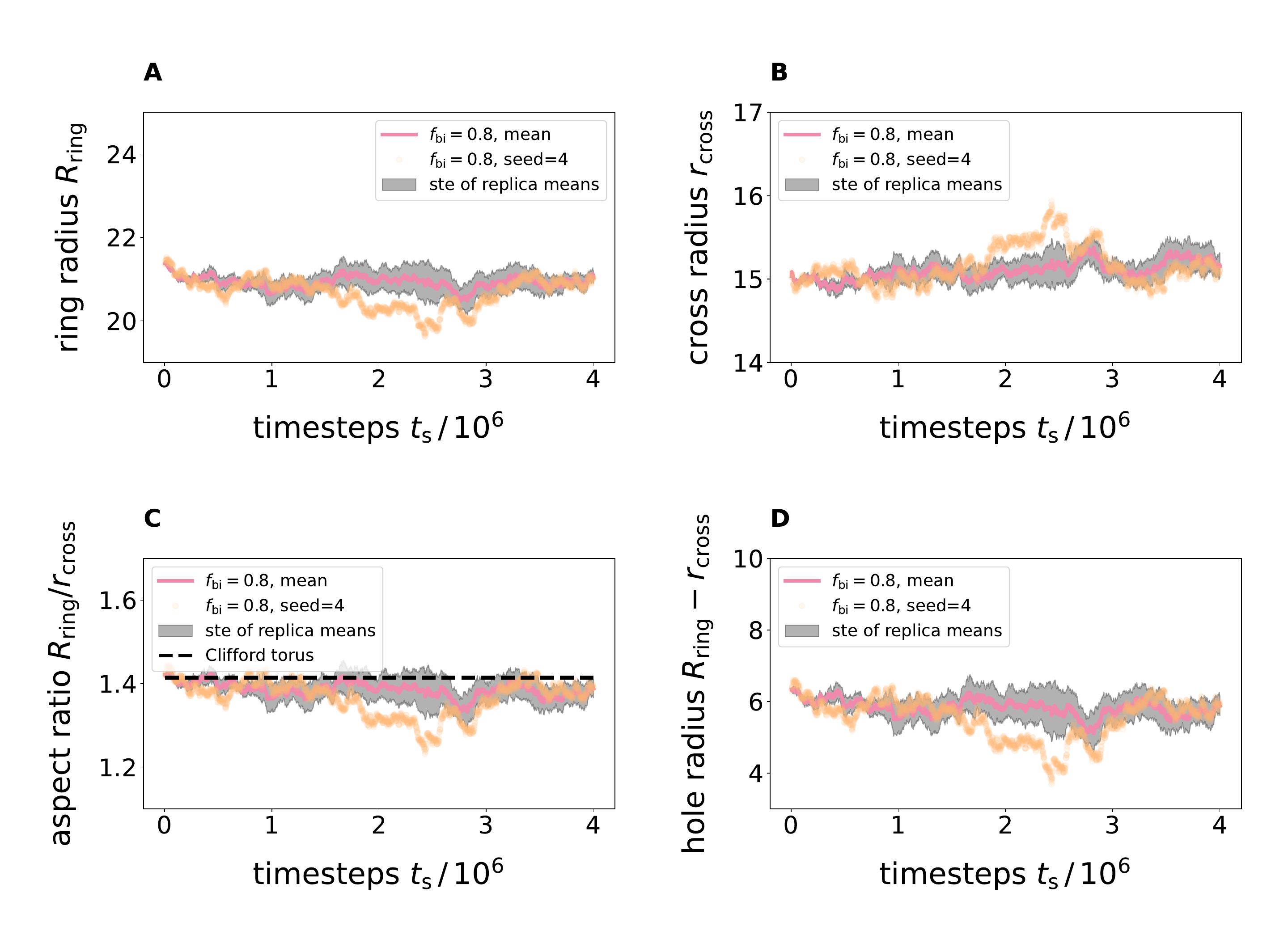}
\caption{Geometrical analysis of toroidal vesicles of mixed membranes at $\fbi=0.8$.
(A) Ring radius as a function of time. 
(B) Cross section radius as a function of time.
(C) Toroidal aspect ratio as a function of time.
(D) Hole radius as a function of time.
Data shown is based on $N_\mathrm{seeds}=4$ random initial seeds.
\label{fig:FigureS5}}
\end{figure}

\begin{figure}[t]
\centering
\includegraphics[width=.95\textwidth]{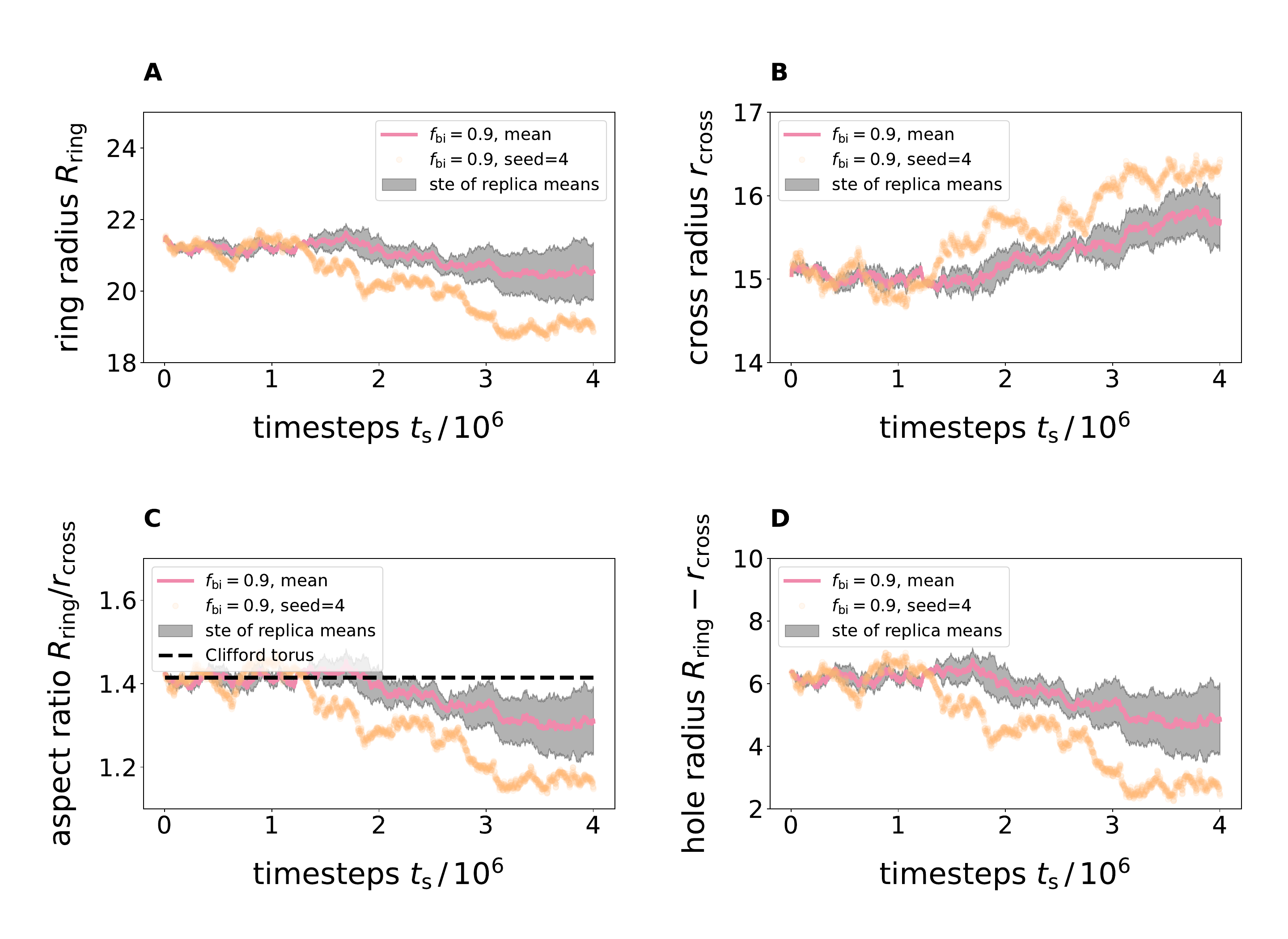}
\caption{Geometrical analysis of toroidal vesicles of mixed membranes at $\fbi=0.9$.
(A) Ring radius as a function of time. 
(B) Cross section radius as a function of time.
(C) Toroidal aspect ratio as a function of time.
(D) Hole radius as a function of time.
Data shown is based on $N_\mathrm{seeds}=4$ random initial seeds.
\label{fig:FigureS6}}
\end{figure}

In Fig.~\ref{fig:FigureS3}, we analyze the geometrical quantities of toroidal vesicles composed of bolalipids at $\km=\unit[0]{}\kBT$.
We show mean values averaged over different seeds, their standard error and a single trajectory.
We find that the ring radius $\Rring$ (A), the cross section radius $\rcross$ (B), the toroidal aspect ratio $\rho=\Rring / \rcross$ (C) and the hole radius $\Rring-\rcross$ (D) change only little over time, implying the torus shape is stable.
In Figs.~\ref{fig:FigureS4}-\ref{fig:FigureS6}, we provide the corresponding analysis for mixture membranes.

To summarize the data, the different quantities are averaged along the mean trajectories starting after $1\cdot10^6$ timesteps until the end of the trajectories.
The results for the toroidal aspect ratios are shown in Table~II.

\subsection{Local fraction of u-shaped bolalipids and bilayer lipids}
\textbf{Local fraction of u-shaped bolalipids}
In this section, we explain how we determine the local fraction of u-shaped bolalipids.
We average over $N_\mathrm{seeds}=16$ trajectories of Clifford tori in steady state with trajectories of $4 \cdot 10^6$ timesteps measured after an initial equilibrium phase of $1 \cdot 10^6$ timesteps (Fig.~4A left).
We extract the position of the membrane lipids every $2\cdot10^3$ simulation steps, corrected for the center of mass position.
Similar to the geometrical analysis, we map the Cartesian coordinates $(x,y,z)$ to spherical coordinates ($R,\theta$) and determine the ring radius $\Rring$ and the cross section radius $\rcross$.
For all membrane lipids, we then determine the toroidal angle $\vartheta=\arctan(R \cos \theta-\Rring /z)$.
The process can be thought of as cutting thin slices of the torus for all azimuthal angles $\varphi$ and stacking them on each other (Fig.~4A right).
We then distribute the lipids in equally populated bins and calculate the u-shaped bolalipid fraction $\uf$ in every bin.
This can be though of as representing the lipid position along a circle with the position parametrized by the toroidal angle $\vartheta$, where $\vartheta=0$ represents the top, $\vartheta=-\pi/2$ the inner shell, $\vartheta=\pi/2$ the outer shell and $\vartheta=\pm \pi$ the bottom (Fig.~4A bottom).
We average $\uf$ and plot the mean value and its error as a function of $\vartheta$ in Fig.~4B.

\begin{figure}[t]
\centering
\includegraphics[width=.95\textwidth]{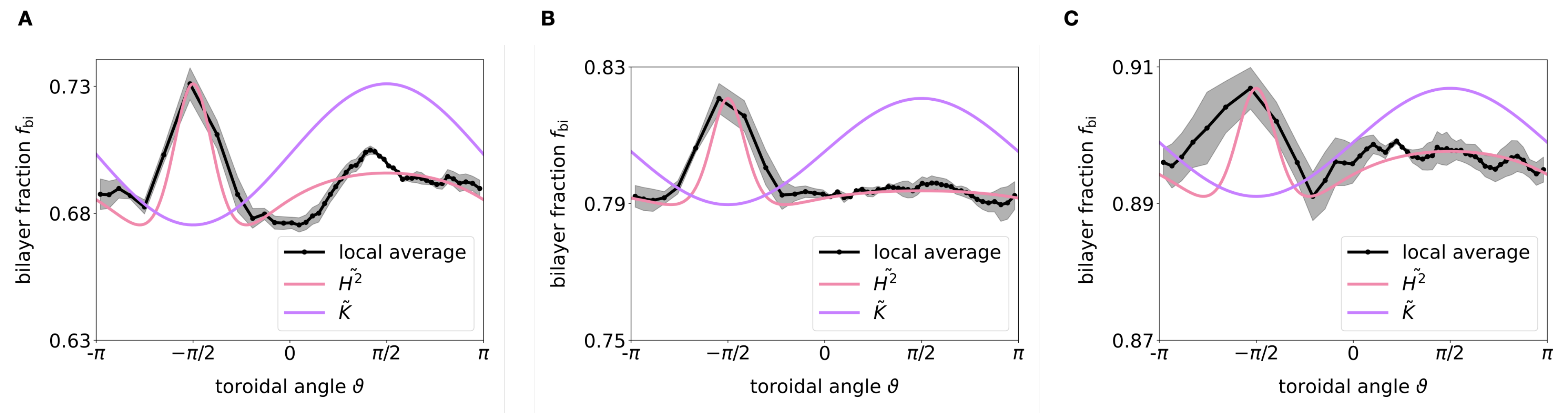}
\caption{The bilayer lipid fraction $\fbi$ as a function of $\vartheta$ (solid black line) with standard error (gray area) for $\fbi=0.7$ (A), $\fbi=0.8$ (B) and $\fbi=0.9$ (C).
Along with $\fbi$, the scaled squared mean curvature $\tilde{H^2}$ (solid red line) and the scaled Gaussian curvature $\tilde{K}$ (solid purple line) are plotted.
Data shown is based on $N_\mathrm{seeds}=4$ random initial seeds.
\label{fig:FigureS7}}
\end{figure}

\textbf{Local fraction of bilayer lipids}
To determine the local fraction of bilayer lipids, we repeat the procedure described above for u-shaped bolalipids.
Instead of u-shaped and straight bolalipids, we now distinguish between bolalipids and bilayer lipids.
$\fbi$ as a function of $\vartheta$ for $\fbi=0.7$ is shown in Fig.~4D and Fig.~\ref{fig:FigureS7}A.
Moreover, $\fbi$ as a function of $\vartheta$ for $\fbi=0.8$ and $\fbi=0.9$  are shown Figs.~\ref{fig:FigureS7}B and C.

\section{Supplemental videos}
\noindent
\textbf{Video S1} Pure bolalipid membranes at different bolalipid rigidities $\km$.
At vanishing bolalipid rigidity ($\km=\unit[0]{}\kBT$), the toroidal vesicle remains stable.
At moderate bolalipid rigidity ($\km=\unit[1]{}\kBT$), the toroidal vesicle transitions smoothly into a spherical vesicle. 
For stiff bolalipids (bolalipid rigidity $\km=\unit[2]{}\kBT$), the toroidal vesicle ruptures.
\\
\\
\textbf{Video S2} Archaeal mixture membranes with varying bilayer lipid fraction $\fbi$. 
At a low bilayer lipid fraction ($\fbi=0.1$), the toroidal vesicle transitions into a spherical vesicle exhibiting transient pores.
At an intermediate bilayer lipid fraction ($\fbi=0.3$), the toroidal vesicle transitions smoothly into a spherical vesicle.
At a large bilayer lipid fraction ($\fbi=0.7$), the toroidal vesicle remains stable. 

\begin{figure}[t]
\centering
\includegraphics[width=.95\textwidth]{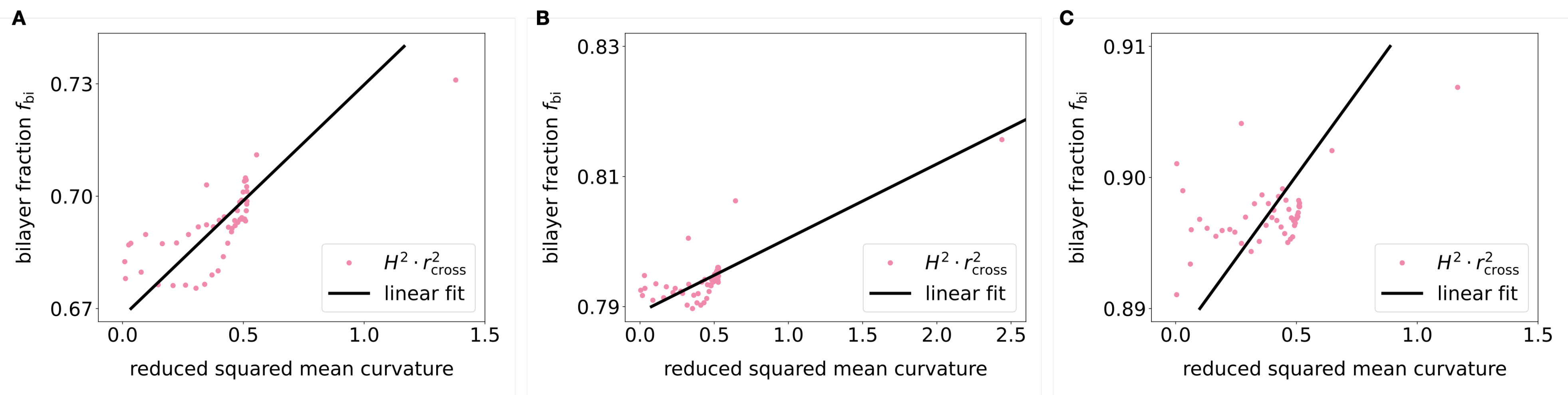}
\caption{The bilayer lipid fraction $\fbi$ as a function of the reduced squared mean curvature $H^2 \cdot \rcross^2$  (red) and a linear fit to the data (solid black line) for $\fbi=0.7$ (A), $\fbi=0.8$ (B), and $\fbi=0.9$ (C).
Data shown is based on $N_\mathrm{seeds}=4$ random initial seeds.
\label{fig:FigureS8}}
\end{figure}

\end{document}